\documentclass[useAMS,usenatbib,psfig]{mn2e} 
\usepackage{graphicx}
\title[Early-type galaxies at large galactocentric radii - II]
{Early-type galaxies at large galactocentric radii - II. \\Metallicity gradients, and the [Z/H]--mass, [$\alpha$/Fe]--mass relations}

\author[Max Spolaor et al.]{Max Spolaor$^1$\thanks{Email: mspolaor@astro.swin.edu.au}, Chiaki Kobayashi$^2$, Duncan A. Forbes$^1$, Warrick J. Couch$^1$, \newauthor George K. T. Hau$^1$\\
$^1$Centre for Astrophysics \& Supercomputing, Swinburne University, Hawthorn, VIC 3122, Australia\\
$^2$Research School of Astronomy and Astrophysics, The Australian National University, Weston ACT 2611, Australia\\}

\begin{document}
\date{Accepted... Received...; in original form 2010}

\pagerange{\pageref{firstpage}--\pageref{lastpage}} \pubyear{2010}

\maketitle
\label{firstpage}

\begin{abstract}
We present the results of a study of stellar population properties at large galactocentric radii of 14 low-mass early-type galaxies in the Fornax and Virgo clusters. 
We derive radial profiles of Age, total metallicity [Z/H], and [$\alpha$/Fe] abundance ratios out to $\sim 1 - 3$ effective radii by using nearly all of the Lick/IDS absorption-line indices in comparison to recent single stellar population models.
We extend our study to higher galaxy mass via a novel literature compilation of 37 early-type galaxies, which provides stellar population properties out to one effective radius. We find that metallicity gradients correlate with galactic mass, and the relationship shows a sharp change in slope at a dynamical mass of $\sim 3.5 \times 10^{10}~\rm M_{\odot}$. 
The central and mean values of the stellar population parameters (measured in $r \leq r_{e}/8$, and at $r  = r_{e}$, respectively) define positive mass trends. 
We suggest that the low metallicities, almost solar [$\alpha$/Fe] ratios and the tight mass-metallicity gradient relation displayed by the low-mass galaxies are indicative of an early star-forming collapse with extended (i.e., $\geq 1$ Gyr), low efficiency star formation, and mass-dependent galactic outflows of metal-enriched gas. The flattening of metallicity gradients in high-mass galaxies, and the broad scatter of the relationship are attributed to merger events.
The high metallicities and supersolar abundances shown by these galaxies imply a rapid, high efficiency star formation. 
The observed [Z/H]--mass and [$\alpha$/Fe]--mass relationships can be interpreted as a natural outcome of an early star-forming collapse.
However, we find that hierarchical galaxy formation models implementing mass-dependent star formation efficiency, varying IMF, energy feedback via AGN, and the effects due to merger-induced starbursts
can also reproduce both our observed relationships.
\end{abstract}

\begin{keywords} galaxies: abundances -- galaxies: dwarf -- galaxies: elliptical and lenticular, cD galaxies: formation -- galaxies: evolution -- galaxies: stellar content
\end{keywords}

\section{Introduction}
The spatial distribution of stellar population properties in early-type galaxies are the chemodynamical fossil imprints of galaxy formation and evolution mechanisms. The efficiency of different physical processes is expected to vary with radius leaving measurable changes at different galactocentric radii. Thereby, providing unique constraints on the galaxy formation mechanisms predicted by competing scenarios.

In the past years an increasing number of observational studies have focused their effort in probing stellar populations beyond the galactic cores (e.g., \citealt{mehlert03}; \citealt{patrizia07}; \citealt{annibali07}). 
However, these previous studies lack to consider low-mass, low-luminosity galaxies because the intrinsic low surface brightness of these objects makes the measurement of their properties at large galactocentric radii observationally challenging. The aim of this paper is to provide a clear picture of the stellar population properties at large radii, and their relationship to galaxy structural parameters over a comprehensive galaxy mass range. 

In the classical models of monolithic collapse (\citealt{larson74a}; \citealt{larson75}; \citealt{carlberg84}; \citealt{arimoto87}; \citealt{pipino04}), early-type galaxies form during dissipational infall of cooling gas. Massive galaxies are predicted to form in a short and intense burst of star formation, whereas dwarf galaxies  experience more extended and intermittent star formation (\citealt{matteucci94}; \citealt{thomas99}; \citealt{chiosi02}). The gas flowing to the centre of the galaxy is chemically enriched by evolving stars, and contributes as metal-rich fuel for star formation. Thus, a negative radial metallicity gradient is established. Metallicity-dependent gas cooling and a time-delay in the occurrence of galactic winds (i.e., onsets that vary with the local escape velocity) can cooperate in steepening any metallicity gradients (\citealt{martinelli98}; \citealt{pipino08}). A strong correlation between metallicity gradient and galaxy mass is expected (\citealt{chiosi02}; \citealt{kawata03}).  [$\alpha$/Fe] abundance gradients are suggested to be a consequence of interplay between local differences in the star formation timescale and gas flows (\citealt{angeletti03}; \citealt{pipino08}).

In the hierarchical merging scenario (e.g., \citealt{cole94}; \citealt{kauffmann98}; \citealt{delucia04}), early-type galaxies form by merging events. The properties of merger remnants are related to the progenitors' mass ratio, amount of gas content and the geometry of the collision (\citealt{mihos94}; \citealt{naab06}). Mergers are expected to flatten any ambient gradients (e.g., \citealt{kobayashi04}). Central bursts of star formation induced by the merger can partially regenerate the metallicity gradients and create age gradients (e.g., \citealt{hopkins09}). A weak relation between mass and metallicity gradients is expected (\citealt{kobayashi04}). Positive or negative [$\alpha$/Fe] gradients can form depending on the chemical enrichment of the pristine gas and the duration of the burst (\citealt{thomasgre99}; \citealt{thokauf99}). Recent models include energy feedback due to active galactic nuclei (AGN) to stop the star formation in massive galaxies, thus producing a mass-dependent star formation history (\citealt{dimatteo05}; \citealt{springel05}; \citealt{croton06}; \citealt{hopkins06}). 

Here we further consolidate the findings of \cite{spolaor09b} (hereafter Paper I) on kinematic and photometric properties at large galactocentric radii of 14 low-mass early-type galaxies in the Virgo and Fornax clusters.
We study the spatial distribution of luminosity$-$weighted Age, total metallicity [Z/H] and $\alpha$-element abundance ratio [$\alpha$/Fe] of out to $\sim 1 - 3 r_{e}$ by using nearly all of the Lick/IDS indices.
The comparison of these line-strength indices with single stellar population models (SSPs) is found to be successful (e.g., \citealt{trager00}; \citealt{proctor02}; \citealt{thomas05}) in breaking the well known age/metallicity degeneracy (\citealt{worthey94}).
We extended our study to a broad galactic mass range, introducing a novel literature compilation of stellar population properties out to one effective radius for a sample of 37 early-type galaxies of higher mass.
We compare properties measured at different galactocentric radii encompassing different percentages of the total stellar mass. The implications for mass scaling relations are considered. By comparing our results to model predictions we constrain the mechanisms for the formation and evolution of early-type galaxies.

This paper is organised as follows. In Section~2, we describe the data sample. In Section~3, we describe our spectroscopic observations and data reduction.
In Section~4, we present the stellar population analysis. In Section~5, we introduce the radial profiles for the 14 low-mass galaxies.
In Section~6, we describe the stellar population properties of the entire galaxy sample. In Section~7, we discuss our results in the context of predictions from competing galaxy formation scenarios. Section~8 presents a summary of our findings.

\section{The data sample}
\subsection{Low-mass galaxies}
Our sample consists of 14 low-mass early-type galaxies in the Virgo and Fornax clusters. 
The six galaxies in the Fornax cluster are chosen from the catalogue of \cite{ferguson89}, while the eight galaxies in the Virgo cluster are from the catalogue of \cite{binggeli85}.
Galaxies of low-mass are selected on the basis on their central stellar velocity dispersion $\sigma_{0}$ and the absolute $B$-band total magnitude $M_{B}$, both of which are known to be independent proxies of galaxy mass. 
The sample uniformly covers the range $1.6 < \log(\sigma_{0}) < 2.15$ and $-16.5 > M_{B} > -19.5$. 
This translates to a dynamical mass range of about $10^{9} < \rm M_{\rm dyn}/ \rm M_{\odot} < 10^{11}$, using the relation $\log($M$_{\rm dyn}) = 2 \log(\sigma_{0}) + \log(r_{e})+ 5.0$ where $r_{e}$ is the effective radius in parsecs (\citealt{cappellari06}).
The main properties of the galaxy sample are summarised in Table~\ref{gal_prop}.

\begin{table*}
\begin{center}
\begin{tabular}{ccccccccccc}
\hline 
\hline 
Galaxy & Alternative & Hubble & $r_{e}$  & P.A.& $\sigma_{0}$ & RV & Distance & $M_{B}$ & M$_{\rm dyn}$ \\
            &     Name    &  Type  & [arcsec] & [degree]&  [km s$^{-1}$] & [km s$^{-1}$] & [Mpc] &[mag] &[10$^{9}$~M$_{\odot}$]\\ 
(1)       &  (2)            & (3) & (4) &(5) &(6) & (7) &(8) &(9) & (10) \\
\hline 
FCC~148 & NGC~1375 & S0 & 14.9  &  89 &    68.9(2.2) &  769.2(5.4) & 18.45 &   -17.30 & 7.9 \\
FCC~153 & IC~1963     & S0 & 12.4  &  81 & 47.8(2.6) & 1635.1(5.6) & 18.45 & -17.10 & 3.1\\
FCC~170 & NGC~1381 & S0 & 13.7 &   139 &   152.9(2.6) & 1749.6(5.9) &18.45 & -18.88 & 34.6\\
FCC~277 & NGC~1428 & E/S0 & 9.8 &   123 &   93.5(2.6) & 1638.3(5.1) & 18.45 & -17.52 & 9.5 \\
FCC~301 & ESO~358-G059 & E/S0 & 10.4 & 156 & 58.9(2.6) & 1040.0(5.3) & 18.45 & -17.02 & 3.9 \\
FCC~335 & ESO~359-G002 & E/S0 & 15.6 & 47 &44.3(2.2) & 1430.8(5.4) & 18.45 & -16.93 & 3.5 \\
VCC~575 & NGC~4318 & E4 & 7.2 &  65  &   88.6(2.6) & 1230.1(5.4) & 22.08 &  -17.61 & 7.2 \\
VCC~828 & NGC~4387 & E5 & 9.1 &   137  & 105.1(2.6) &  589.1(5.3) & 17.94 &  -18.26 & 10.7\\
VCC~1025 & NGC~4434 & E0/S0 & 12.9 & 134 &135.0(2.6) & 1070.5(5.1) & 22.39 &  -18.72 & 31.6\\
VCC~1146 & NGC~4458 & E1 & 26.1 & 175  & 102.1(2.6) &  677.3(5.7) & 16.37 & -18.13 & 26.9\\
VCC~1178 & NGC~4464 & E3 & 6.6 &  14  & 108.1(2.6) & 1266.2(5.2) &	15.84 &  -17.63 & 7.2\\
VCC~1297 & NGC~4486B & E1 & 2.3 &  110 & 273.8(2.6) & 1570.1(5.3) & 16.29 &  -16.77 & 16.2\\
VCC~1475 & NGC~4515 & E2 & 9.4 &   15  & 91.4(2.6) &  951.0(5.8) & 16.59 &  -17.84 & 7.7\\
VCC~1630 & NGC~4551 & E2 & 12.5 &  67  & 110.4(2.6) & 1181.5(5.4) & 16.14 &  -18.06 & 1.0 \\
\hline \hline
\end{tabular}
\end{center}
\caption[]{ Galaxy properties. Col.(1): galaxy name from the catalogue
 of \cite{ferguson89} and \cite{binggeli85}; (2): alternative galaxy name; (3): morphological type from the HyperLEDA database; (4): $B$-band effective radius along the galaxy semi-major axis from the HyperLEDA database; (5): position angle of the major axis; (6, 7): central stellar velocity dispersion and radial velocity from Paper~II; (8): distance to the galaxy using the distance modulus derived from surface brightness fluctuations of \cite{tonry01}, \cite{mei05} and \cite{blakeslee09}; we applied the correction of \cite{jensen03} to the values of Tonry; (9): total $B$-band absolute magnitudes estimated from total apparent magnitudes given in the NASA/IPAC Extra-galactic Database and corrected for Galactic extinction; (10): dynamical mass of the galaxy using the $\log($M$_{\rm dyn}) = 2 \log(\sigma_{0}) + \log(r_{e})+ 5.0$, where $r_{e}$ is the effective radius expressed in parsecs (\citealt{cappellari06}).}
\label{gal_prop}
\end{table*}

\subsection{High-mass galaxies}
\label{high_mass}
We extend our study to higher galaxy mass by using data for 37 early-type galaxies compiled from the literature.
The properties of these galaxies and the specific studies they have been drawn from are given in Table~\ref{gal2_prop}.

We stress that the observing and data reduction procedures used in these literature studies closely resemble those used in this work.
The final sample of 51 early-type galaxies is uniquely uniform due to the use of nearly all of the Lick/IDS indices and the steps taken in the stellar population analysis.
Specifically, the same $\chi^{2}$ minimization technique and single stellar population models have been used to convert Lick/IDS indices to stellar population parameters (see Sec.~4 for more details).
The data sample comprehensively covers the range $1.6 < \log(\sigma_{0}) < 2.6$ and $-16.5 > M_{B} > -23.5$, corresponding to $2.5 \times 10^{9} <~$M$_{\rm dyn}/$M$_{\odot} < 4 \times 10^{11}$. 

For completeness, we briefly describe the six studies from which these data are obtained.
\citet{spolaor08a, spolaor08b} studied the star formation and chemical evolutionary history of the two brightest galaxies of the NGC~1407 group. 
\cite{patrizia07} used Keck LRIS long-slit spectra to study a sample of 11 early-type galaxies covering a wide range in mass and environment.
\cite{brough07}  focused on the stellar population and kinematic properties of three nearby brightest group galaxies and three brightest cluster galaxies located at z~$\sim$~0.055.
\cite{reda07} analysed the metallicity gradients of 12 isolated luminous early-type galaxies. 
The sample of \cite{proctor03} consists of 11 early-type galaxies located in the Leo cloud and the Virgo cluster. 
We note that five of the 11 galaxies of the \cite{patrizia07} sample overlap with the \cite{proctor03} sample.
Since their stellar population parameters are similar we decide to use the values from \cite{proctor03} because they were derived from all 25 Lick/IDS indices with respect to the 20 used by \cite{patrizia07}.

\section{Observations and data reduction}
\subsection{GMOS long-slit spectroscopy}
The observations were made using the Gemini Multi-Object Spectrograph (GMOS) mounted on the Gemini South telescope in long-slit spectroscopy mode. The six galaxies in the Fornax cluster were observed in semester 2006B (Program ID GS-2006B-Q-74), while the eight galaxies in the Virgo cluster were observed in semester 2008A (Program ID GS-2008A-Q-3).

Four 1800s exposures were taken for each galaxy except FCC~335, for which two exposures of 1800s were taken.  At the beginning of each exposure, the slit was centred on the nucleus and oriented along the major axis of the galaxy. The GMOS detector comprises three 2048~$\times$~4608~EEV~CCDs with a pixel size of 13.5~$\times$13.5~$\mu$m$^{2}$. We adopted the standard full frame readout and, to enhance the signal-to-noise, we selected to bin 2$\times$~spatially, yielding a resolution of 0.145~arcsec~pixel$^{-1}$. 

The slit width was set to 1.0~arcsec for the 2006B run and 0.5~arcsec for the 2008A run. It was 5.5$\arcmin$ in length in both cases. For the 2006B run, we used the 600 line~mm$^{-1}$ grism blazed at 5075~\AA,~providing a dispersion of 0.457~\AA~pixel$^{-1}$
and a wavelength coverage of 3670$-$6500~\AA. For the 2008A run, we used the 600 line~mm$^{-1}$ grism blazed at 5250~\AA~with 0.458~\AA~pixel$^{-1}$, covering the wavelength range of 3860$-$6670~\AA. The instrumental set up provided a spectral resolution, as measured from the full width at half maximum~(FWHM) of the arc lines, of $\sim$~4.21~\AA~(2006B run) and $\sim$~2.45~\AA~(2008A run). During the observations, the seeing varied from 0.7-arcsec to 1.0-arcsec FWHM, which is more than adequate for the purposes of our study. Spectrophotometric and radial velocity standard stars covering a range of spectral classes were observed as calibrators. These stars belong to both the Lick/IDS stellar library (\citealt{worthey94}) and MILES library (\citealt{patrizia06}). Bias frames, dome flat-fields and copper$-$argon (CuAr) arcs were also taken for calibration. 

\subsection{Data reduction}
The data reduction was carried out using the Gemini/GMOS IRAF package (version 1.9.1). The software also computes the data quality and variance planes, producing an associated error spectrum for each data spectrum. This allows us to perform an accurate error analysis, with full propagation of errors through to the derivation of stellar population parameters.

Each object frame is treated separately. Initial reduction of the CCD frames involves overscan correction, bias and dark subtraction, flatfield correction and cosmic ray removal. The flatfield correction is performed by means of both a quartz-halogen lamp (taken with the Gemini calibration unit, GCAL) and twilight spectra, which are normalised and divided from all the spectra. This process corrects for pixel-to-pixel sensitivity variations and for two-dimensional large-scale illumination patterns due to slit vignetting. Cosmic rays are identified and eliminated by interpolating each line of an image with a high order function. The residuals (cosmic ray hits) that are narrower than the expect instrumental line width are replaced with the fitted value. The package GSWAVELENGTH is used to establish an accurate wavelength calibration from copper$-$argon (CuAr) arc images taken just before and after each exposure. Each spectrum is converted to a linear wavelength scale using roughly 130 arc lines fitted by 3th-5th order polynomials. An accuracy of better than 0.1~\AA~is consistently achieved for the wavelength calibration solution along the whole spectral coverage. The spectra are also corrected for geometrical distortions along the spectral directions (S-distortion). The spectral resolution is derived as the mean of the Gaussian FWHMs measured for a number of unblended arc-lamp lines which are distributed over the whole spectral range of a wavelength-calibrated spectrum. 

To obtain reliable measurements of the Lick/IDS indices at large galactic radii, it is critical to perform an accurate sky subtraction. Sky contamination becomes increasingly significant for the outer parts of the galaxy. In these regions, the level of galaxy light can be a few percent of the sky signal. \cite{cardiel95} show the effect of incorrect sky subtraction on the measurement of line-strength indices radial profiles. This effect represents one of the main sources of systematic errors in the analysis of the stellar population radial trends.
The generous field-of-view of GMOS and the intrinsic low luminosity of our sample allow us to accurately estimate the sky continuum from the edges of the slit, where the galaxy light is negligible. The sky level is estimated by interpolating along the outermost 10$-$20 arcsec at the two edges of the slit and subtracted from the whole image. A sky subtraction accurate to within 1 percent is achieved, guaranteeing negligible related systematics on the measured indices.

For each galaxy, we obtain four fully reduced galaxy frames of identical exposure time. These are then co-added to form a single frame. The final galaxy frame is a two-dimensional spectrum from which we extract 1D spectra along the slit. The spatial width (i.e.~the number of CCD rows binned) for each extracted 1D spectrum increases with radius to achieve a minimum signal-to-noise (S/N) per \AA~of $\sim$~35 in the spectral region of Mg$_{b}$. By adopting this criterion we have been able to obtain reliable measurements of line-strength indices and, therefore, stellar population radial trends to galactocentric radii of $\sim 3 r_{e}$.

\section{Stellar population analysis}
The wavelength range of our spectroscopic observations covers the 21 line-strength indices of the original Lick system defined in \cite{faber85} and \cite{worthey94}, plus the 4 additional age-sensitive Balmer lines H$\delta_{A}$, H$\gamma_{A}$, H$\delta_{F}$, and H$\gamma_{F}$ as presented in \cite{worthey97}. The raw line-strength indices are measured using the method described in \cite{proctor02} and adopting the definition of \cite{trager98} for index spectral pseudocontinua and bandpasses. The raw indices are then calibrated to the Lick/IDS system and converted to stellar population parameters of luminosity-weighted Age, total metallicity [Z/H] and $\alpha$-element abundance ratios [$\alpha$/Fe] by comparison with single stellar population models. The details of this procedure are described in the following sub-sections.

\subsection{Calibrations to the Lick/IDS system}
The procedure to calibrate raw indices to the Lick/IDS system involves three different steps: (i) the correction for differences in the spectral resolution between the Lick/IDS system and our observations; (ii) the correction for broadening due to internal velocity dispersion of the observed galaxies; (iii)  the correction for differences in flux calibration between our data and the stellar calibration spectra due to continuum shape offsets. The procedure is based on the method presented in \cite{proctor02}, but see also Appendix~A in \cite{spolaor08b}.

Briefly, the spectra of galaxies with resolution higher than that of the Lick/IDS system are convolved with an appropriate Gaussian prior to index measurement. The Lick/IDS spectral resolution for each index is estimated from fig.~7 of \cite{worthey97}, as in table~2 of \cite{proctor02}. When the opposite condition is found and the resolution exceeds that of the Lick/IDS system, the indices are measured from the observed Lick/IDS standard stars after convolving the spectra with a series of Gaussians of known widths. This procedure allows us to estimate the appropriate correction factors for indices in our galaxies. In order to estimate the differences in flux calibration we measure the indices in each of the observed standard stars after broadening to the appropriate resolution. The difference between our measurements and those taken on the Lick/IDS system (\citealt{worthey94}) of the same standard stars are given in Table~\ref{offsets}. The measured offsets are then applied to the spectral resolution and velocity dispersion broadening corrected indices. Index errors arise from two types of errors which are added in quadrature: (i) random statistical errors due to Poisson noise; (ii) uncertainties in recession velocity, velocity dispersion and systematic errors originating from the conversion onto the Lick/IDS system quantified using the error on the mean of the offsets. The systematic and random errors are of the same order (e.g, \citealt{proctor04}; \citealt{patrizia07}). These errors are taken into account in the single stellar population model fitting procedure.

\begin{table}
\begin{center}
\begin{tabular}{cccc}
\hline
\hline
Index & Unit & Offset & Error in Mean\\
\hline
H$\delta_{A}$ & \AA &0.143 & 0.182\\
H$\delta_{F}$ & \AA &-0.225 & 0.200\\
CN$_{1}$ & mag &0.004  &  0.010 \\
CN$_{2}$ & mag & 0.011  &  0.010 \\
Ca4227 & \AA &0.209  &  0.100\\
G4300 & \AA &0.071  &  0.230\\
H$\gamma_{A}$&\AA& -0.156 & 0.074\\
H$\gamma_{F}$&\AA&-0.115 & 0.152\\
Fe4383  & \AA  & -0.033 & 0.125\\
Ca4455  &  \AA &-0.054  &   0.053\\
Fe4531  & \AA  &0.346     &  0.103\\
C4668   & \AA  &-0.335    &  0.187\\
H$_{\beta}$& \AA &-0.021 &    0.083\\
Fe5015  & \AA & -0.013  &  0.158\\
Mg$_{1}$ & mag& 0.027  &  0.010\\
Mg$_{2}$ & mag& 0.029  &  0.010\\
Mg$_{b}$ & \AA& 0.088   & 0.072\\
Fe5270   & \AA& 0.000   & 0.140\\
Fe5335  & \AA& 0.015   & 0.042\\
Fe5406 &  \AA& 0.100  &  0.185\\
Fe5709 & \AA&  0.031  &  0.065\\
Fe5782 & \AA&  0.026   &  0.067\\
NaD    & \AA& -0.239  &   0.138\\
TiO$_{1}$ & mag& 0.006   &  0.003\\
TiO$_{2}$ & mag&  -0.015  &  0.005  \\
\hline 
\hline
\end{tabular}
\end{center}
\caption{Offsets for the observed standard Lick/IDS stars. Estimates are between our index measurements and those taken from the original Lick/IDS system (\citealt{worthey97}).}
\label{offsets}
\end{table}

\subsection{Emission-line correction}
Emission lines of different intensity can be seen in a large percentage of early-type galaxies. 
They affect the measurement of the H$\beta$, H$\gamma_{A,F}$, H$\delta_{A,F}$, Fe5015, and Mg$_{b}$ Lick/IDS indices.

In order to estimate the flux from emission lines we use the GANDALF software (\citealt{sarzi06}). The software finds the combination of template spectra which best reproduce the galaxy spectrum and simultaneously fits the emission lines as additional Gaussian templates. Thus, the emission spectra are subtracted from the galaxy spectra and the Lick/IDS indices are measured in the emission-free spectra. We use template spectra from the MILES standard star library (\citealt{patrizia06a}). The approximate sensitivity to the detection of emission lines in a typical 1D spectrum of our dataset, which has a median S/N~$\sim$~35, is $\sim$~0.27~\AA. 
We find that only few galaxies present emission lines, and their intensity level is weak.

\subsection{Stellar population model-fitting}
We compare the measured Lick/IDS indices to single stellar population models of \citet{thomas03, thomas04}. 
The advantage of these models is that they reproduce the effects of varying $\alpha$-abundance ratios on Lick/IDS indices.
For each index, a 3-dimensional grid of model $\log$(Age), [Z/H], and [$\alpha$/Fe] values are provided. The grid encompasses values for $-1 < \log \rm(Age) < 1.175$ dex, and $-2.25 < \rm [Z/H] < 0.8$ dex in steps of 0.025 dex, and $-0.3 < [\rm \alpha/Fe] < 0.6$ dex in 0.03 dex steps.

The best fit to the stellar population parameters $\log \rm(Age)$, [Z/H] and [$\alpha$/Fe], is achieved by simultaneously fitting the 3-dimensional model grids throughout a $\chi^{2}$-minimisation of as many observed indices as possible. The technique reduces the $\chi^{2}$ deviation between the observed and the modelled index values taking into account the index errors (\citealt{proctor02}). The strength of the method is that it works with as many indices as possible in order to break the age-metallicity degeneracy that affects each index differently. 
Observed indices that are highly deviant from model values can be easily flagged and excluded by the fitting process through a 3$\sigma$ clipping procedure.
The fit is re-iterated on the remaining indices until a stable solution is found. 
We stress that this methodology allows us to derive values of stellar population parameters that are robust against data reduction uncertainties such as flux calibration errors, stray cosmic rays, sky-line residuals, velocity dispersion errors, errors in the conversion to the Lick system, emission line-filling, etc.

The errors on the derived stellar population parameters are estimated by means of Monte Carlo simulations. The best-fitting SSP model values are convolved with a Gaussian of width equal to the index error estimates. Hence, stellar population parameters are measured from index values which are randomly selected from the simulated distribution. The final errors are the rms values derived by repeating this procedure 50 times. We find that median errors estimated from our observational errors are $\pm$~0.07~dex,  $\pm$~0.04~dex, and $\pm$~0.05~dex  respectively for $\log \rm (Age)$, [Z/H], and [$\alpha$/Fe]. The uncertainties associated with the specific SSP models produce systematic errors larger than those estimated by the Monte Carlo realisations.  Although the model-related errors become statistically significant when comparing our results to those of other works that use different SSP models, they do not affect the analysis of stellar population radial gradients or relationships within our data sample.

\subsection{Radial profiles of Ages, Metallicities, and Abundance Ratios}
\begin{figure*}
\includegraphics[scale=.7]{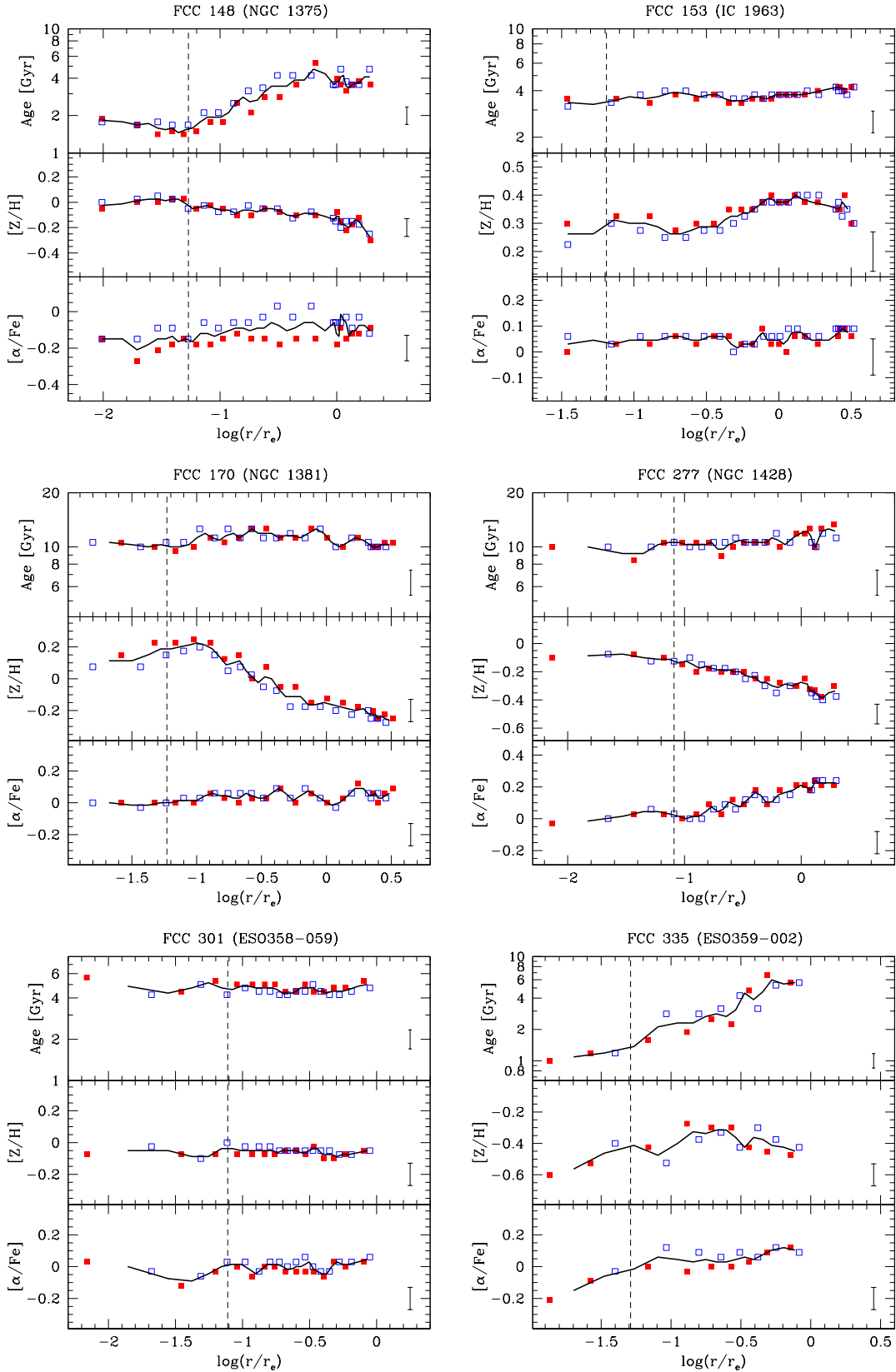}
\caption[]{The major-axis stellar population parameters of the Fornax cluster low-mass galaxies. Values are estimated from spatially resolved apertures extracted along the negative (open blue squares) and positive (filled red squares) side of the major axis. The points are folded onto the positive side. The x-axis indicates the galactocentric distance of each extracted aperture, scaled with the galaxy's effective radius and expressed in logarithmic scale. The vertical dashed black line represents the seeing limit. The solid black line is a weighted average of the data points. The error bars express the median error in the derived parameters. }
\label{pop_fornax}
\end{figure*}

The radial profiles of Age, [Z/H], and [$\alpha$/Fe] measured along the major axis of the 14 low-mass galaxies are presented in Fig.~\ref{pop_fornax}, Fig.~\ref{pop1_virgo}, and Fig.~\ref{pop2_virgo}. The radial profiles extend out to $\sim 1- 3 r_{e}$.

Values are expressed as a function of the galactocentric radius of the aperture from which they are derived. 
This radius is then scaled with the effective radius to allow us a consistent comparison between profiles of different galaxies.
The profiles have been folded onto the positive side of the major axis with respect to the galaxy's centre. In this way, we can evaluate asymmetries between the two sides.
We find that parameters measured at same radii from the two major axis's sides are similar within the errors, suggesting a large-scale chemical coherence.

In Appendix~\ref{individual} we give a description of the observed radial profiles for each galaxy in the sample.

\begin{figure*}
\includegraphics[scale=.7]{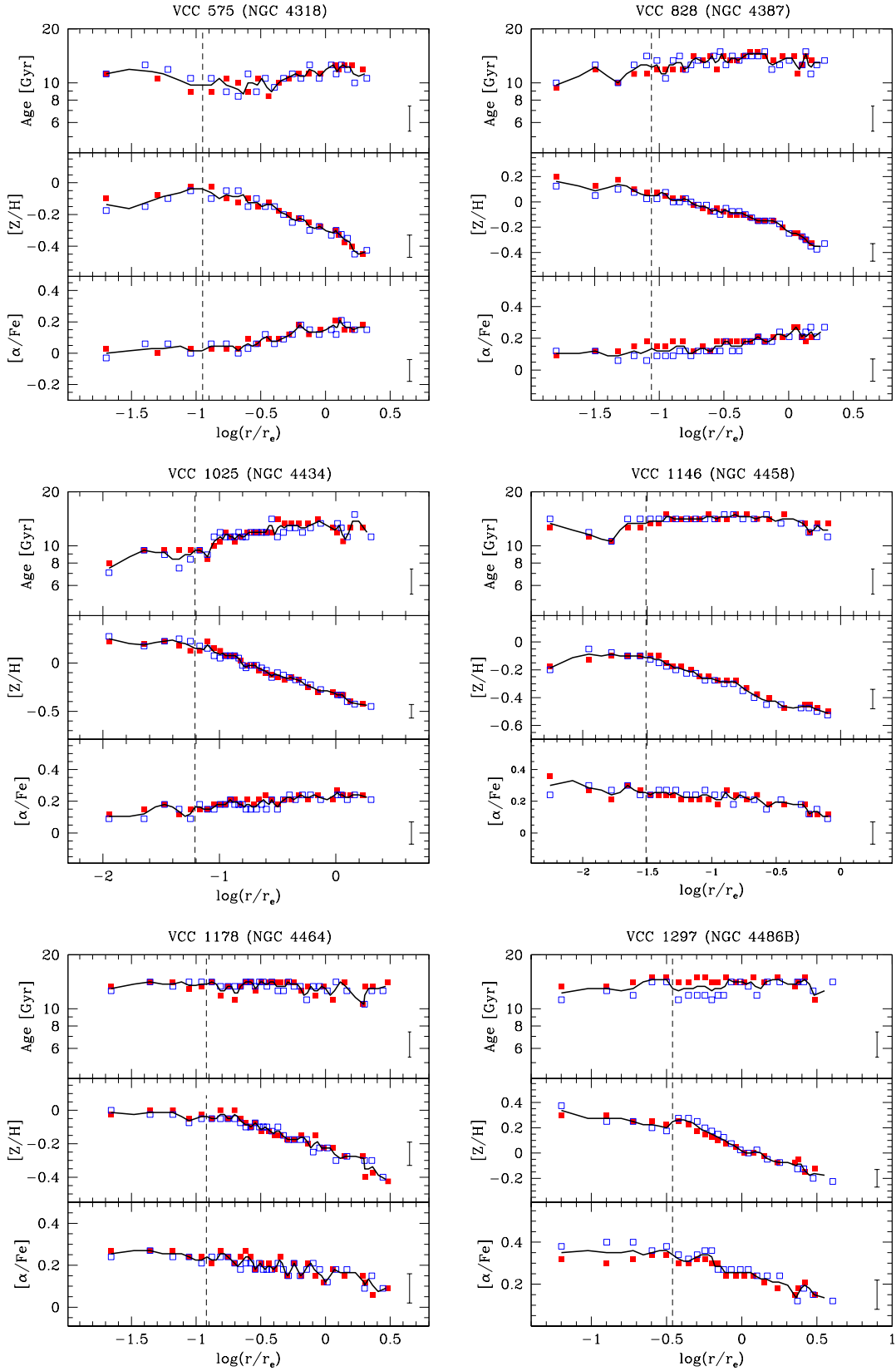}
\caption[]{The major-axis stellar population parameters of the Virgo cluster low-mass galaxies. Description is as for Fig.~\ref{pop_fornax}.}
\label{pop1_virgo}
\end{figure*}

\begin{figure*}
\includegraphics[scale=.7]{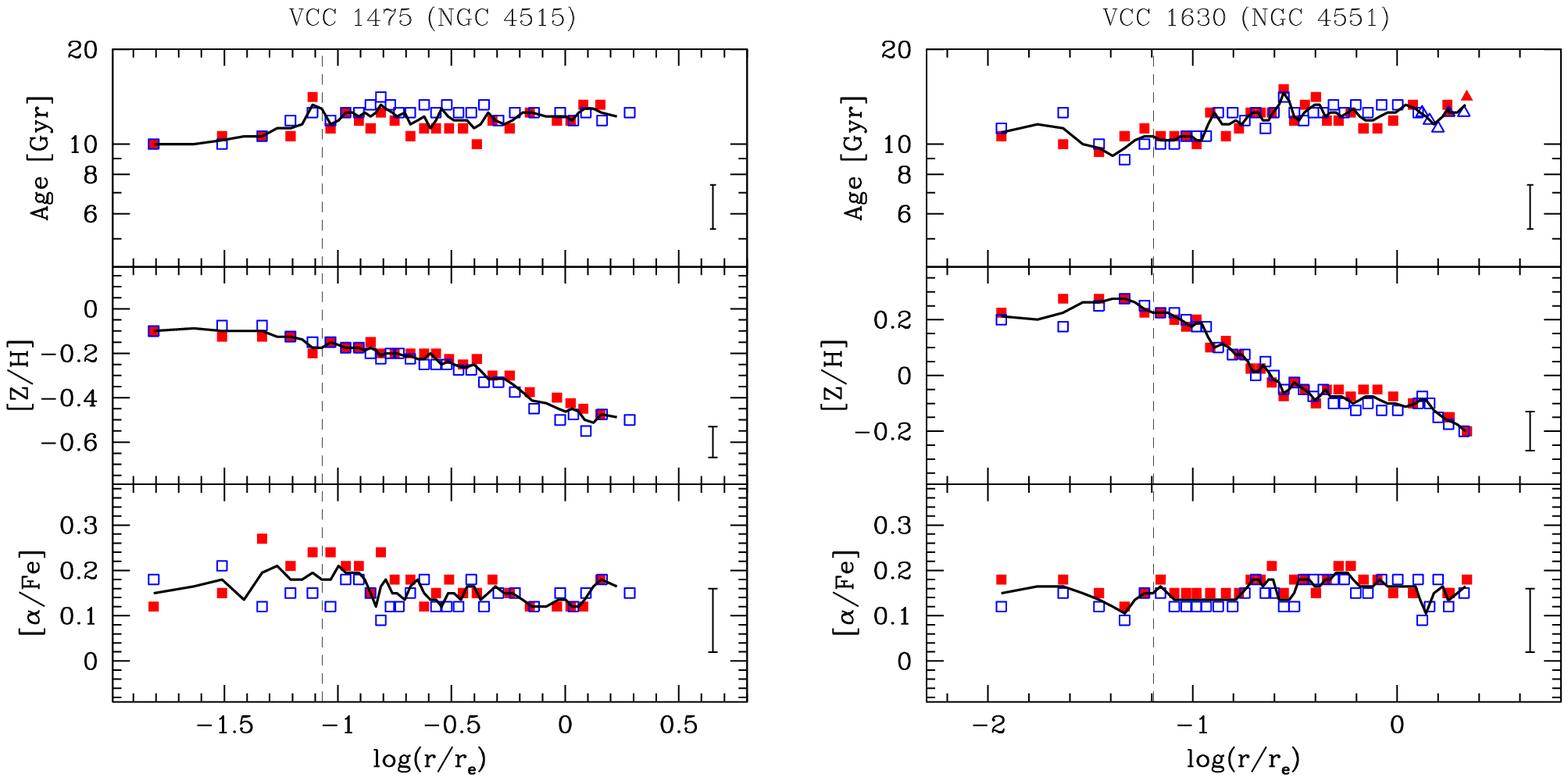}
\caption[]{Continued from Fig.~\ref{pop1_virgo}.}
\label{pop2_virgo}
\end{figure*}

\section{Galaxy properties}
In this section, we combine our results on low-mass galaxies with the properties of the high-mass galaxy sample described in Section~\ref{high_mass}. This allows us to analyse stellar population properties at large radii over a comprehensive mass range.

\subsection{Stellar population gradients}
We measure radial gradients of stellar population by performing a weighted linear least-squares fit to the data points, of the form:
\begin{equation}
\rm P(\textit{r}) = P_{\textit{r}_{e}} + \frac{\Delta P}{\Delta \log \textit{r}} \log\frac{\textit{r}}{\textit{r}_{e}} \; ,
\end{equation}
where $r$ is the galactocentric radius, $r_{e}$ the effective radius and P$_{r_{e}}$ is the stellar population parameter evaluated at $r_{e}$.
Gradients express the derivative variation of each stellar population parameter along the galactocentric radius scaled by the effective radius.
We denote them as grad~Age, grad~[Z/H] and grad~[$\alpha$/Fe].
The fit considers data points out to one effective radius, but it excludes points from apertures extracted within a radius equal to the seeing disc.
By doing this we can consistently compare our results with gradients measured in the data sample of high-mass galaxies. We report the gradient values in Table~\ref{Pop_par1}.

We find that gradient values remain statistically similar when we include points beyond one effective radius, i.e. out to the largest observed radii.
We have verified this to be valid for all the low-mass galaxies in our sample with extended stellar population radial profiles.
However, we can not confirm that this is the case for the high-mass galaxies, due to fact that their stellar population profiles do not extend beyond one effective radius.

\begin{figure*}
\includegraphics[scale=0.70]{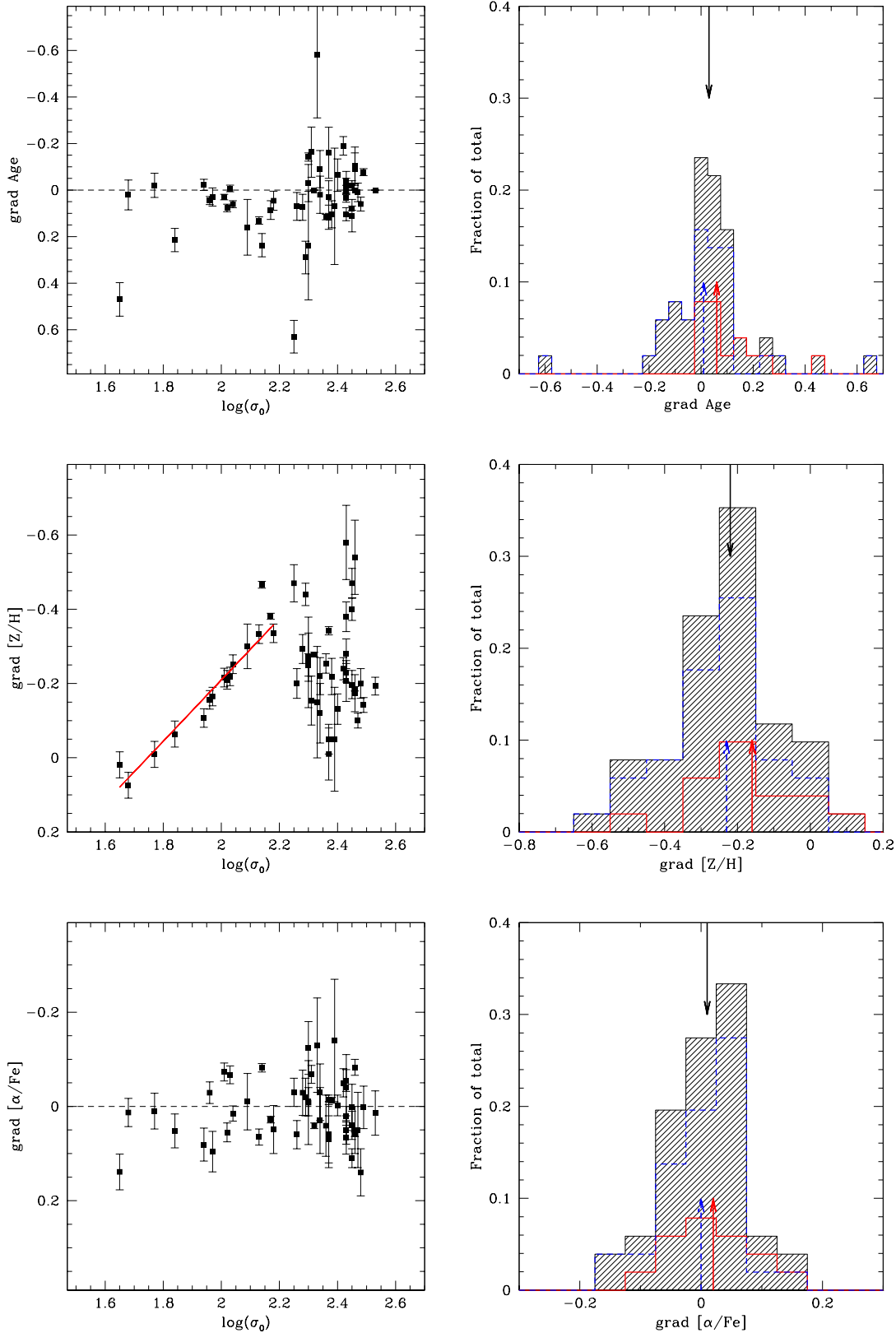}
\caption[]{Stellar population gradients of our sample galaxies. Left plots (from top to bottom):  age, metallicity and abundance ratio gradients as a function of central stellar velocity dispersion (a proxy for galaxy mass). In the grad~[Z/H]--$\log(\sigma_{0})$ diagram, the red line shows a weighted least-squares linear fit to the data points for low-mass galaxies. Right plots (from top to bottom):
histograms of the stellar population gradients. In each plot, the distributions for the entire data sample (black shaded region), the low-mass galaxies (red solid line) and the high-mass galaxies (blue dashed lines) are shown. The median values are shown as superimposed arrows: at the top of the histograms for the entire data sample (black arrow) and at the bottom for low-mass galaxies (red solid arrow) and high-mass galaxies (blue dashed arrow).}
\label{grads}
\end{figure*}

In Fig.~\ref{grads}, we show the relation between stellar population gradients and central velocity dispersion $\sigma_{0}$ of galaxies.
The central velocity dispersion is known as galaxy mass indicator (e.g., \citealt{graves09}) and it is measured  by averaging the dispersion values within a radius of $r_{e}/8$.

We find that metallicity gradients of early-type galaxies correlate well with galactic mass.
We find no significant gradients in Age and [$\alpha$/Fe] and no relation against the mass.

Originally reported in \cite{spolaor09a}, the mass--metallicity gradient relation shows a different trend for low-mass (i.e., $1.6 < \log(\sigma_{0}) \leq 2.15$) and high-mass (i.e., $2.15 < \log(\sigma_{0}) < 2.6$) galaxies, with a sharp transition point at $\log(\sigma_{0}) \sim 2.15$. The trend established by low-mass galaxies is remarkably tight, and their metallicity gradients become less negative and hence shallower with decreasing mass. At the very low-mass end, $\log(\sigma_{0}) < 1.8$, gradient values change sign becoming positive or nil. In the range $1.6 <\log(\sigma_{0}) < 2.15$, the slope of the relation is $-0.83$~$\pm$~$0.05$.
The transition point at $\log(\sigma_{0}) \sim 2.15$ marks the beginning of a downturn from the tight low--mass trend. 
Galaxies with increasing mass are characterised by shallower metallicity gradients. 
However, the downturn is marked by broad scatter and some of the most massive galaxies have steeper negative gradients. 
A non-parametric Spearman rank-order test, performed on all data points in the range $2.15 < \log(\sigma_{0}) < 2.6$, gives a level of significance lower than 0.04; that is, there is a probability greater than 96 per cent that a correlation exists.

The number distribution of stellar population gradients for the entire data sample are shown in Fig.~\ref{grads}. Median values are given in Table~\ref{stats}.
We also plot histograms of the subsamples of low-mass galaxies, $1.6 < \log(\sigma_{0}) \leq 2.15$, and high-mass galaxies, $2.15 < \log(\sigma_{0}) < 2.6$.

The metallicity gradient distributions peak at negative gradient values. 
The entire data sample and the high-mass galaxies have similar distributions, with almost equal median values.
On the other hand, low-mass galaxies have on average shallower negative gradients.
In the histograms of Age and [$\alpha$/Fe] gradients for the entire data sample and the high-mass galaxies the peaks are at almost null gradient value.
Low-mass galaxies have on average very shallow positive Age and [$\alpha$/Fe] gradients.

\begin{table}
\begin{center}
\begin{tabular}{cccc}
\hline 
\hline 
Median & Low-mass   & High-mass & Entire Sample \\
		&	galaxies & galaxies &\\
\hline
grad~Age & 0.06 $\pm$ 0.13 & 0.01 $\pm$ 0.18 & 0.03 $\pm$ 0.17\\
grad~[Z/H] & $-$0.16 $\pm$ 0.15 & $-$0.23 $\pm$ 0.13 & $-$0.22 $\pm$ 0.14\\
grad~[$\alpha$/Fe] & 0.02 $\pm$ 0.07 & 0.00 $\pm$ 0.06 & 0.01 $\pm$ 0.06\\
\hline
Age$_{0}$  [Gyr]  & 10.0 $\pm$ 2.1& 10.5 $\pm$ 1.6& 10.2 $\pm$ 1.8\\
$ \rm [Z/H]_{0}$   & 0.00 $\pm$ 0.22&0.32 $\pm$ 0.12&0.29 $\pm$ 0.20\\
$ \rm [\alpha/Fe]_{0}$ &0.11 $\pm$ 0.13 &0.27 $\pm$ 0.11&0.24 $\pm$ 0.14\\
\hline
$\langle  \rm Age \rangle$ [Gyr]& 11.7 $\pm$ 1.5& 11.7 $\pm$ 1.5 & 11.7 $\pm$ 1.5\\
$\langle  \rm [Z/H] \rangle$ & $-$0.27 $\pm$ 0.22 &$-$0.05 $\pm$ 0.20 &$-$0.13 $\pm$ 0.22\\
$\langle  \rm [\alpha/Fe] \rangle$ & 0.15 $\pm$ 0.11 &0.25 $\pm$ 0.12 &0.22 $\pm$ 0.13\\
\hline 
\hline
\end{tabular}
\end{center}
\caption[]{Statistics of stellar population properties for our sample galaxies; stellar population gradients, central and mean stellar values.
The median values are for distributions of low-mass galaxies ($1.6 < \log(\sigma_{0}) \leq 2.15$), high-mass galaxies ($2.15 < \log(\sigma_{0}) < 2.6$), and the entire data sample ($1.6 < \log(\sigma_{0}) < 2.6$).}
\label{stats}
\end{table}

Our results are in agreement with previous findings of stellar population gradients in high-mass galaxies.
\cite{mehlert03}, \cite{annibali07} and \cite{rawle08} considered stellar population gradients in early-type galaxies out to almost one effective radius.
They analysed objects with central velocity dispersion in the range $2.1 \leq \log(\sigma_{0}) \leq 2.5$. 
In general, Age and [$\alpha$/Fe] gradients are found to be negligible, whereas metallicity gradients become shallower with increasing mass.
The median values derived for the metallicity gradients distributions in these studies are respectively $-0.16$, $-0.21$ and $-0.20$ dex per decade.
The lower value found by \cite{mehlert03} is probably due to the fact that they derive gradients of total metallicity [Z/H] from radial profiles of just a few spectral indices Mg, Fe and H$\beta$. 
In \cite{kobayashi99}, the average of metallicity gradients was as steep as $-0.3$ dex per decade, and no relation between gradients and mass is found.

\subsection{Central and mean stellar population properties}
\label{comparison}
Central values of stellar population parameters Age$_{0}$, [Z/H]$_{0}$, [$\alpha$/Fe]$_{0}$ are calculated by averaging the values from all apertures extracted within a radius of $r_{e}/8$. The parameters for all sample galaxies are reported in Table~\ref{Pop_par2}. 

Following \cite{kobayashi99}, we define mean values of stellar age $\langle \rm Age \rangle$, total metallicity $\langle \rm [Z/H] \rangle$ and $\alpha$$-$abundance ratio $\langle \rm [\alpha/Fe] \rangle$ of a galaxy as the parameters measured at one effective radius. The difference between this definition and the average of values from apertures within one effective radius is related to the stellar population gradients, and is negligibly small (see Eq.(22) of \citealt{kobayashi99}).
We find that if the gradient absolute value is smaller than 0.3 dex per decade then the two mean values differ of $\sim$~0.05~dex.
If the gradient's absolute value ranges between 0.3 and 0.6 dex per decade then a difference of $\sim$ 0.09~dex is detected.
These values are comparable to the stellar population parameters errors estimated by our stellar population model-fitting analysis.
Mean stellar values of our sample galaxies are given in Table~\ref{Pop_par3}.

\begin{figure*}
\includegraphics[scale=0.63]{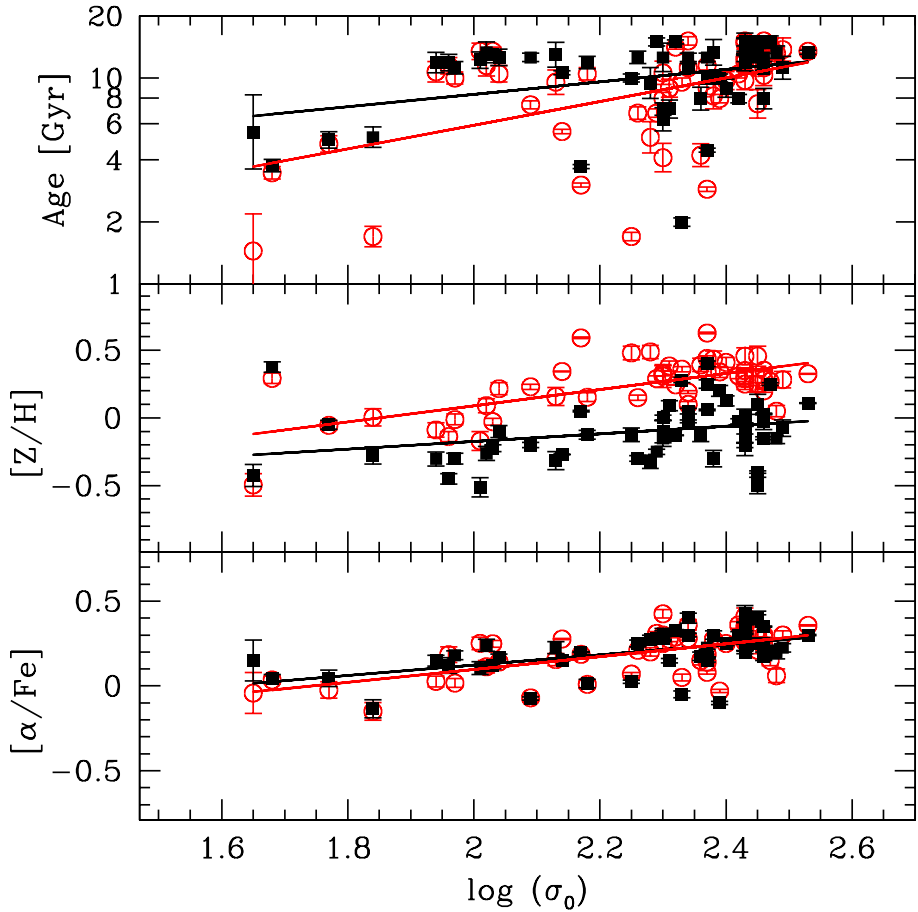}
\caption[]{Stellar population parameters as a function of central stellar velocity dispersion (as proxy for galaxy mass). Open red circles and filled black squares respectively indicate central values and mean stellar values of stellar population parameters of our sample galaxies.
The red and black solid lines are the weighted least-squares linear fit to the data points.}
\label{overplot_wfits}
\end{figure*}

\begin{figure*}
\includegraphics[scale=0.55,angle=-90]{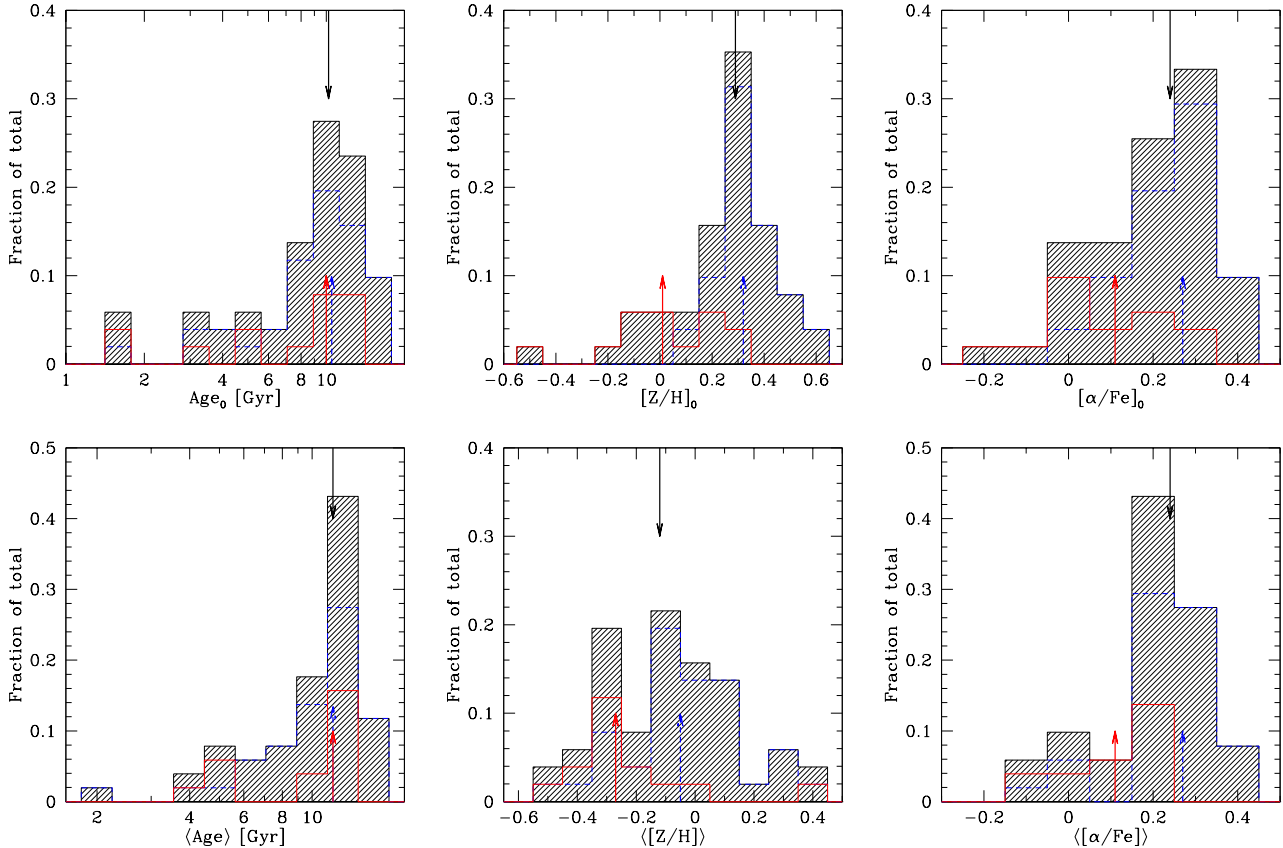}
\caption[]{Histograms of central and mean values of stellar population parameters.  In each plot, the distributions for the entire data sample (black shaded region), the low-mass galaxies (red solid line) and the high-mass galaxies (blue dashed lines) are shown. The median values are shown as superimposed arrows: at the top of the histograms for the entire data sample (black arrow) and at the bottom for low-mass galaxies (red solid arrow) and high-mass galaxies (blue dashed arrow).}
\label{overplot}
\end{figure*}

In Fig.~\ref{overplot_wfits}, we compare the mass trends defined by central and mean stellar population parameters.
In Fig.~\ref{overplot} we plot the histograms of central and mean stellar parameters for the entire data sample and for the subsamples of low-mass ($1.6 < \log(\sigma_{0}) \leq 2.15$) and high-mass galaxies ($2.15 < \log(\sigma_{0}) < 2.6$). Median values are given in Table~\ref{stats}. 

\begin{table*}
\begin{center}
\begin{tabular}{ccccc}
\hline 
\hline 
 Relation&  & Age--$\log(\sigma_{0})$ & [Z/H]--$\log(\sigma_{0})$ & [$\alpha$/Fe]--$\log(\sigma_{0})$\\
\hline 
Central stellar values &  Slope           & $0.57 \pm 0.14$  & $0.59 \pm 0.10$  & $0.38 \pm 0.07$ \\
Mean stellar values  &  & $0.30 \pm 0.11$ & $0.28 \pm 0.13$  & $0.30 \pm 0.07$ \\
\hline
Central stellar values& Zero point&  $-0.38 \pm 0.33$ &  $-1.10 \pm 0.23$ & $-0.66 \pm 0.16$\\
Mean stellar values&   & $0.32 \pm 0.26$ & $-0.74 \pm 0.30$ & $-0.49 \pm 0.16$\\
\hline \hline
\end{tabular}
\end{center}
\caption[]{Best-fitting parameters to our central and mean values of stellar population parameters versus mass relations.}
\label{gal_stat2}
\end{table*}

We find a clear mass-metallicity relation for our sample galaxies, such that the metallicity increases with increasing galactic mass.
The comparison between the distributions of [Z/H]$_{0}$ and $\langle \rm [Z/H] \rangle$ values indicates that central galactic regions are more metal-rich then the rest of the galaxy. We find that our best-fitting parameters, given in Table~\ref{gal_stat2}, of the central and mean metallicity-mass relations reveal a statistically 
significant difference not only in zero point but also in slope. 
This is attributed respectively to stellar population metallicity gradients and to the mass-metallicity gradient relation.
In fact, we have previously found that the sample galaxies have an average metallicity gradient of $-0.23 \pm 0.14$ dex per decade.
However, we have also found that metallicity gradients in low-mass galaxies are on average flatter than those in high-mass objects.
This effect is such that the [Z/H]$_{0}$--$\log(\sigma_{0})$ and $\langle \rm [Z/H] \rangle$--$\log(\sigma_{0})$ fits tend to converge with decreasing mass.
The residuals to the best fit are found not to correlate with other parameters.

The mass--[$\alpha$/Fe] relations for our data sample has a remarkably small scatter that is preserved along the entire mass range.
Massive galaxies have significantly supersolar abundance ratios, while low-mass galaxies have ratio comparable to stars in the solar neighbourhood.
The mass--$\langle \rm [\alpha/Fe] \rangle$ and mass--[$\alpha$/Fe]$_{0}$ relationships are statistically equivalent (Table~\ref{gal_stat2}).
A similar conclusion is reached by inspecting the distributions of the two abundance ratio parameters.

The age of the galaxies is found to be the least variable among the three plotted parameters.
In general, galaxies of similar mass have ages which can differ by several gigayears.
However, there exists a weak trend such that most massive galaxies have older ages than low-mass systems.
We performed a weighted least squares fit to the data points; best-fitting parameters are reported in Table~\ref{gal_stat2}.
The mean stellar age of galaxies is observed to be older than their central age. 
In fact, the Age$_{0}$ and $\langle \rm Age \rangle$ distributions peak at old age values and their median values differ by $\sim$~2 Gyrs.

In Table~\ref{gal_stat}, we summarise the findings by previous studies on mass trends of central stellar population values.
A consistent comparison is hindered by differences in the data samples (i.e., mass range), and in the techniques adopted in estimating stellar population parameters. 
However, we find that the work of \cite{nelan05} is the most comparable to our study.
In particular, they derive central stellar population properties of 4097 red-sequence galaxies from the NOAO Fundamental Plane Survey (NFPS), by comparing all of the Lick/IDS indices to SSP models of \citet{thomas03,thomas04}. 

In the literature, the existence of an Age$_{0}$--$\log(\sigma_{0})$ relation is still under debate. 
\cite{jorgensen99}, \cite{kuntschner01}, \cite{mehlert03}, and \cite{patrizia06} (high-density environment galaxies) report a lack of trend in their samples, whereas \cite{poggianti01}, \cite{caldwell03}, \cite{thomas05}, \cite{nelan05}, \cite{bernardi06}, \cite{smith06}, \cite{patrizia06} (low-density environment galaxies), \cite{annibali07} and \cite{graves07} claim a clear correlation between the two parameters. 
\cite{nelan05} claim a steepen of the relation for galaxies with central stellar velocity dispersion values less than $\log(\sigma_{0}) \sim 2.1$. Our results do not allow us to confirm this change in slope, but we observe a sudden drop to younger age values for low-mass galaxies in comparison to the older high-mass objects.
All previous studies find a correlation between central metallicity and mass of early-type galaxies. 
However, the measured slope values can differ up to 30 percent.  Our results are in good agreement with the values of \cite{nelan05}.
The slope values for the [$\alpha$/Fe]$_{0}$--$\log(\sigma_{0})$ relation measured by \cite{mehlert03}, \cite{thomas05}, \cite{nelan05}, \cite{bernardi06}, \cite{smith06}, \cite{annibali07} and \cite{graves07} are similar to ours.

\begin{table*}
\begin{center}
\begin{tabular}{ccccc}
\hline 
\hline 
Reference & Range in $\log(\sigma_{0})$  & Age$_{0}$--$\log(\sigma_{0})$  & [Z/H]$_{0}$--$\log(\sigma_{0})$ & [$\alpha$/Fe]$_{0}$--$\log(\sigma_{0})$\\
\hline 
\cite{kobayashi99} & 2.1 $-$ 2.6 & $-$ & 0.76 & $-$\\
\cite{kuntschner01} & 2.0 $-$ 2.6   & none & 0.9 & $-$ \\
\cite{mehlert03} & 1.9 $-$ 2.6 & none  & 0.77 $\pm$ 0.10 & 0.36 $\pm$ 0.09 \\
\cite{thomas05}  & 2.1 $-$ 2.5  & 0.78 $\pm$ 0.23  & 0.42 $\pm$ 0.14 & 0.36 $\pm$ 0.05\\
\cite{nelan05} & 1.9 $-$ 2.4 & 0.59 $\pm$ 0.13  & 0.53 $\pm$ 0.08 & 0.31 $\pm$ 0.06\\
\cite{patrizia06} & 2.1 $-$ 2.5 & weak/none$^{\dag}$  & 0.00 $\pm$ 0.00 & $-$\\
\cite{bernardi06} & 2.0 $-$ 2.5 & 0.81  & 0.58 & 0.39\\
\cite{smith06} & 1.9 $-$ 2.4 & 0.72 $\pm$ 0.14 & 0.37 $\pm$ 0.08 & 0.35 $\pm$ 0.07\\
\cite{clemens06} & 2.0 $-$ 2.5 & $-$ & 0.76 & 0.74\\
\cite{annibali07} & 2.1 $-$ 2.5 & 0.43  & 0.48 & 0.41\\
\cite{graves07} & 2.0 $-$ 2.4 & 0.35 $\pm$ 0.03 & 0.79 $\pm$ 0.05 & 0.36 $\pm$0.04\\
\bf{This work} & 1.6 $-$ 2.6   & 0.57 $\pm$ 0.14  & 0.59 $\pm$ 0.10  & 0.38 $\pm$ 0.07  \\
\hline 
\hline
\end{tabular}
\end{center}
\caption[]{Literature compilation of best-fitting slopes to central values of stellar population parameters versus mass relations. $^{\dag}$: \cite{patrizia06} find a slope of 0.00177~$\pm$0.00033 for low-density environment galaxies, whereas they don't find a relationship for galaxies in high-density environments.}
\label{gal_stat}
\end{table*}

\subsection{Relations between stellar population parameters}
\begin{figure}
\includegraphics[scale=0.43]{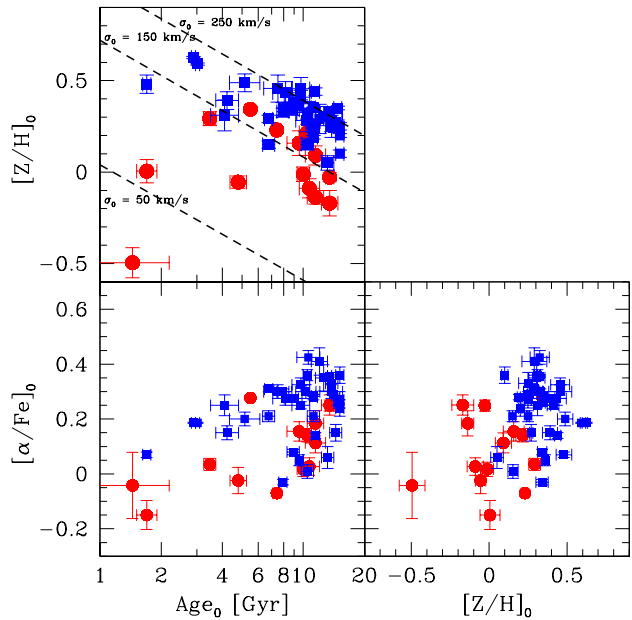}
\caption[]{ Relations between central values of stellar population parameters. Dashed lines indicate constant velocity dispersion in the metallicity hyperplane. Red filled circles are low-mass galaxies. Blue filled squares are high-mass galaxies.}
\label{par1}
\end{figure}

The relationships between central stellar population parameters Age$_{0}$, [Z/H]$_{0}$, and [$\alpha$/Fe]$_{0}$ of our sample are shown in Fig.~\ref{par1}. These parameters are known to individually correlate with mass (e.g.,~\citealt{trager00}). Therefore, in the plots we encode the data points as low-mass and high-mass objects, divided at a transitional mass of $\log(\sigma_{0}) = 2.15$.

We find that galaxies with younger central age have a higher central metal content, which implies that the young stars are formed out of pre-enriched gas.
\cite{jorgensen99}, \cite{poggianti01} and \cite{thomas05} observe a similar trend in their data sample of early-type galaxies, although \cite{kuntschner01} argues that the observed age-metallicity anticorrelation is more likely to be an artefact caused by errors due to the well known age-metallicity degeneracy (\citealt{worthey94}). 

\cite{trager00}, \cite{trager08} and \cite{smith08} find that Age$_{0}$, [Z/H]$_{0}$, and [$\alpha$/Fe]$_{0}$ together with $\log(\sigma_{0})$ of galaxies form a four-dimensional space and their variation can be described by two separate bidimensional relationships: (i) a correlation between [$\alpha$/Fe] and $\log(\sigma_{0})$; (ii) the metallicity hyperplane. The hyperplane is described by a linear combination of Age$_{0}$, [Z/H]$_{0}$, and $\log(\sigma_{0})$ of the form:

\begin{equation}
\rm [Z/H] = \alpha \log(\sigma_{0}) + \beta \log Age + \gamma
\end{equation}

\noindent The best-fitting parameters of a weighted least squares fit to our data samples are $\alpha = 0.80 \pm 0.08$ , $\beta = 0.36 \pm 0.09$  and $\gamma = 1.24 \pm 0.04$. Therefore, the Age$_{0}$--[Z/H]$_{0}$ diagram in Fig.~\ref{par1} is a projection of the metallicity hyperplane.
Galaxies with old central ages tend to have significantly larger $[\alpha/Fe]_{0}$ values than younger objects.
This result is often interpreted as [$\alpha$/Fe] being an indicator for star formation timescales. 
Finally, massive galaxies have higher [Z/H]$_{0}$ and [$\alpha$/Fe]$_{0}$ values with respect to more metal-poor and lower abundance ratio low-mas galaxies.

The mean values of stellar population parameters $\langle \rm Age \rangle$, $\langle \rm [Z/H] \rangle$, and $\langle \rm [\alpha/Fe] \rangle$ are found not to correlate with each other.
In particular, the trends observed for central values are erased when mean stellar values are considered. 

Finally, we systematically surveyed all parameter combinations for possible correlations with stellar population gradients, however we did not detect any significant correlations.

\section{Discussion}
\subsection{Origin of metallicity gradients}
\begin{figure}
\includegraphics[scale=0.43]{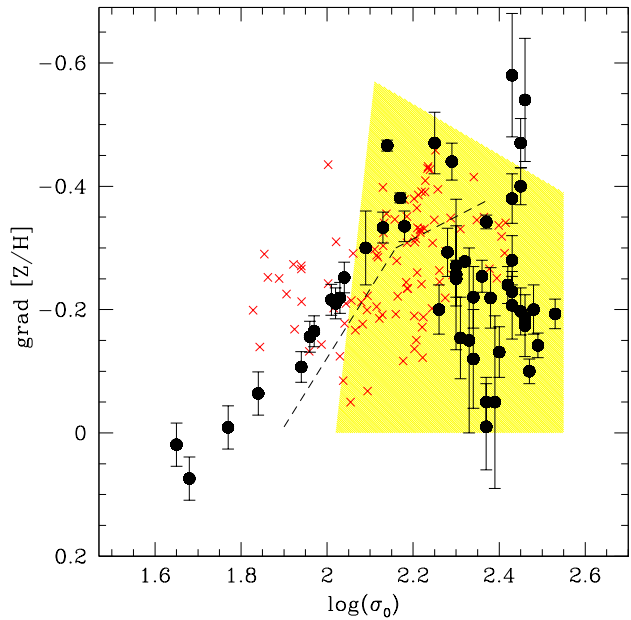}
\caption[]{Metallicity gradients as a function of galaxy central velocity dispersion. Black points indicate our observations. 
The predictions of an early star-forming collapse of \cite{kawata03} are shown as black dashed lines.
The region occupied by the remnants of major mergers between gas-rich disc galaxies, as simulated by \cite{hopkins09}, is shown by the light yellow shading. 
The red crosses indicate merger remnants simulated by \cite{kobayashi04}.}
\label{for_paper}
\end{figure}

Below we discuss the origin of metallicity gradients in our sample galaxies in the context of competing galaxy formation mechanisms. In Fig.~\ref{for_paper}, the mass-metallicity gradient relation is compared to numerical simulations by \cite{kawata03}, \citet{kobayashi04,kobayashi05} and \cite{hopkins09}.

The chemodynamical simulations of \cite{kawata03} model the formation of early-type galaxies via the dissipative collapse of a gas cloud, upon which CDM perturbations are superimposed. 
\citet{kobayashi04,kobayashi05} simulate the formation and chemodynamical evolution of field elliptical galaxies with a variety of merging histories in a CDM-based scenario.
A number of physical processes, such as radiative cooling, star formation, energy feedback via Type II and Ia supernovae and stellar winds, and chemical enrichment from intermediate mass-stars are included in the models. \cite{hopkins09} reproduce elliptical galaxies with a variety of metallicity gradients in simulations of equal-mass mergers of gas-rich disc galaxies, including feedback processes from galactic winds and black hole growth. 

The strong grad~[Z/H]--mass trend exhibited by our low-mass galaxies (i.e., $1.6 < \log(\sigma_{0}) \leq 2.15$) is in good agreement with the predictions of \cite{kawata03}. 
In order to produce the steepening of gradients with increasing mass, an early star-forming collapse with varying star formation efficiency is required. The interplay between the efficiency of radiative cooling and feedback heating is such that the gas infall rate decreases with increasing galactic mass. In more massive galaxies, the sinking gas is chemically enriched with a greater efficiency. Consequently, the rate at which new metal-rich stars are formed and therefore the final slope of the metallicity gradient increase with increasing mass. The almost solar [$\alpha$/Fe] abundance ratios found for the galaxies imply a gradual conversion of gas into stars. The tight [$\alpha$/Fe]--mass relation that they form indicates a star formation efficiency that varies with mass (see also next section).

In Paper I, we studied the kinematic and photometric properties of these galaxies, and the results support the above scenario.
The galaxies are found to host old stellar discs which extend to the largest observed radii (i..e, $\sim 3 r_{e}$). This cold stellar component is the dominant source of their dynamical support.
Gas dissipation and redistribution of angular momentum during an early star-forming collapse are suggested to be responsible for generating the observed large amount of rotational support.
In particular, low star formation efficiency with a gradual conversion of gas into stars at all galactic radii (i.e., out to $\sim$ 3$r_{e}$) is required to explain the observed prevalence of discy-shaped orbits and the radial variation of their ellipticity. 

The metallicity gradients in the galaxies simulated by \citet{kobayashi04,kobayashi05} are steeper at $\log(\sigma_{0}) < 2$, and with a larger scatter around $\log(\sigma_{0}) \sim 2.2$ than those of our low-mass galaxies (i.e., $1.6 < \log(\sigma_{0}) \leq 2.15$; see Fig.~\ref{for_paper}). In the simulations, the large scatter originates from the difference in the merging histories, with the flatter gradients formed by major merger. 
Although we need further data to confirm it, the small scatter observed in our relationship could indicate an homogeneous formation history, in agreement with our previous findings of an early star-forming collapse, and that intermediate-mass ($\log(\sigma_{0}) \sim 2.2$) early-type galaxies are not formed by major merger.

The sharp trend transition at $\log(\sigma_{0}) \sim 2.15$ in the grad~[Z/H]--$\log(\sigma_{0})$ diagram (Fig.~\ref{for_paper}) is such that for galaxies of higher mass (i.e., $2.15 < \log(\sigma_{0}) < 2.6$) the relationship deviates from the predictions of \cite{kawata03}, and shows a large scatter.
In the literature, this mass value (i.e., $3.5 \times 10^{10} \rm M_{\odot}$) often marks a transition in the properties of physical mechanisms acting in early-type galaxies (e.g., \citealt{kauffmann03}; \citealt{croton06}; \citealt{cattaneo08}).
The gradients in high-mass galaxies tend to flatten with increasing mass producing a downturn from the low-mass regime.
The merger-based models of \cite{hopkins09} well reproduce the region occupied by metallicity gradients of high-mass galaxies.

The principal factors that contribute in flattening the gradients are the progenitors' mass ratio and their initial pre-merger gradients.
For example, progenitors with identical initial gradients and a gas content that is 10 percent of the total stellar mass of the discs, produce a final gradient that is almost 50 percent shallower than that generated by progenitors with a gas content of 40 percent. In a merger between progenitors with identical initial gradients but without merger-induced star formation, the remnant has a gradient up to 40 percent flatter than the originals (\citealt{dimatteo09}).
Only 3~Gyr after the merging, time evolution is predicted to weaken the gradients up to 10 percent of their original slope.
For completeness, we note that the kinematic and photometric properties of more massive galaxies are well known to be consistent with the predictions of gas-rich disc galaxies mergers (e.g. \citealt{bender94}; \citealt{naab06}; \citealt{cox06}). We attribute the downturn to be a consequence of merging and the broad scatter a natural result of the different merger properties.

The situation for the massive galaxies with the steeper negative gradients is more puzzling. These galaxies are classified as brightest group and cluster galaxies and their gradients are not reproduced by the models. A possible interpretation is that after their merger assembly, their position at the bottom of the gravitational potential well of group/cluster of galaxies favours the accretion of surrounding gas from cooling flows. This induces further central star formation thus steepening the metallicity gradients. However, the old stellar age (i.e., $\geq 8$ Gyrs) of the central regions place the last major gas-accretion event at redshift $z\geq 1$. Alternatively, these galaxies were formed in an early dissipational collapse during a very efficient star-forming episode. The high efficiency in converting gas into stars coupled with energy feedback processes prompted the steep gradients. Consequent passive evolution preserves the gradients.

\subsection{Do the [Z/H]-mass and [$\alpha$/Fe]-mass relations have a common origin?}

The physical mechanism endorsed by chemodynamical models (e.g., \citealt{matteucci94}; \citealt{kawata03}; \citealt{kobayashi04}) and cosmological hydrodynamic simulations (e.g., \citealt{delucia04}; \citealt{brooks07}; \citealt{derossi07}; \citealt{finlator08}; \citealt{tassis08}), for explaining the observed [Z/H]--mass, and [$\alpha$/Fe]--mass relationships is a mass-dependent star formation efficiency.

In this interpretation, the conversion efficiency of gas-phase metals into new stars decreases to lower galaxy mass, leading to extended star formation.
The low metallicities observed in our low-mass galaxies could indicate a depletion of metal-enriched gas by mass-dependent galactic outflows, due to energy feedback (i.e., mass-dependent galactic winds), and facilitated by the shallow gravitational potential well.
On the other hand, high-mass galaxies are able to retain a greater amount of the chemical enriched gas, which therefore can be converted and locked into new metal-rich stars.

The  [$\alpha$/Fe] ratio can be used to distinguish between rapid, efficient star formation and extended, less efficient star formation.
This is motivated by the time-delay in the production of Fe-peak elements with respect to $\alpha$-elements, due to the different timescales of Type Ia and Type II SNe.
An analytical link between [$\alpha$/Fe] ratio and star formation timescale was proposed by \cite{thomas05}.
The [$\alpha$/Fe] is proportional to the logarithm of the timescale of a star-forming episode, assuming a Gaussian star formation rate (SFR).
They show that a composite stellar population with a ratio [$\alpha$/Fe]~=~0.2 requires a formation timescale $\leq$ 1 Gyr.
The supersolar abundance ratio observed in the high-mass galaxies implies a rapid and efficient star formation episode.
The star formation history of our low-mass galaxies is expected to last for several gigayears, thus yielding solar [$\alpha$/Fe] ratio, and is required to be less efficient.
In particular, a long period of time is necessary for the metal contribution of Type Ia SNe to become relevant such that it dilutes the [$\alpha$/Fe] ratio.

The almost negligible [$\alpha$/Fe] gradients displayed by our galaxies, at any given galaxy mass and out to large radii, might indicate that regions at different galactocentric radius have similar star formation timescales.

The above given interpretation of the observed [Z/H]--mass [$\alpha$/Fe]--mass relationships is a natural outcome of the early star-forming collapse scenario.
On the other hand, the failure in simultaneously reproducing the [Z/H]--mass and [$\alpha$/Fe]--mass relationships is a well known problem for models of the hierarchical merging scenario (e.g., \citealt{thomasgre99}).
The complication lies in the fact that mechanisms required to satisfy the former relation tend to worsen the agreement with the latter, and vice-versa (e.g., \citealt{pipino08c}).
In particular, the [$\alpha$/Fe]--mass relation is often interpreted as observational evidence for downsizing (e.g., \citealt{somerville08}; \citealt{trager09}; \citealt{fontanot09}).

\cite{thomas99} and \cite{thokauf99} were the first to study the [$\alpha$/Fe] abundance ratio in models with star formation histories derived from a hierarchical semi-analytical galaxy formation model. The models of \citet{nagashima05a,nagashima05b} included the chemical enrichment from a variety of elements due to Type Ia and Type II SNe and a varying initial mass function. However, all of these studies obtain a decrease of the [$\alpha$/Fe] abundance ratio with increasing mass because of the over-extended star formation histories in high-mass galaxies. In the simulations of \cite{pipino08b}, a significant improvement at high masses is obtained by quenching the star formation via AGN feedback in massive haloes. They suggest that a better suppression of star formation at low and intermediate masses is required to avoid the presence at high redshift of low-mass supersolar $\alpha$-elements enhanced galaxies.
This is an essential requirement to hierarchically build more massive galaxies with the observed [$\alpha$/Fe] abundance ratios.

\begin{figure}
\includegraphics[scale=0.43]{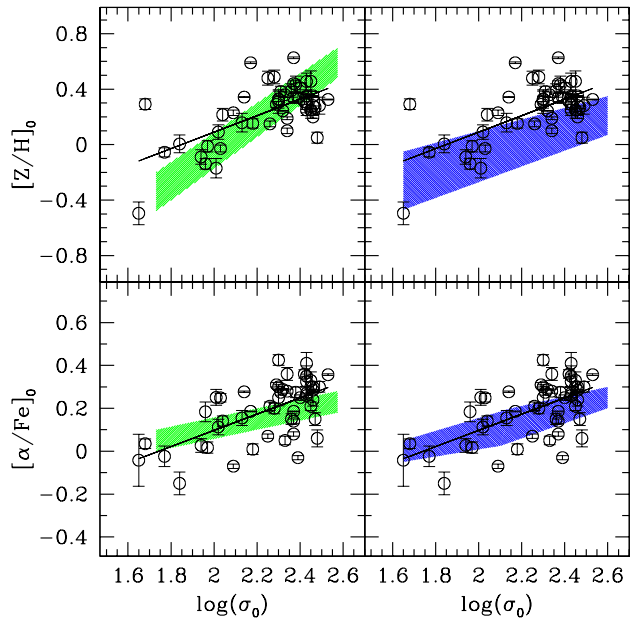}
\caption[]{Comparison between observed and predicted [Z/H]$_{0}$--mass and [$\alpha$/Fe]$_{0}$--mass relationships. Open black circles represent central values of our sample galaxies. The black solid line is the weighted least-squares linear fit to the data points.
In the left plots, we plot the predictions from \cite{arrigoni09} (green shading). In the right plots, the models of \cite{calura09} are shown (blue shading).}
\label{overplot_wfits1}
\end{figure}

Recently, the simulations of \cite{arrigoni09} and \cite{calura09b} coupled detailed galactic chemodynamical evolution models to semi-analytic hierarchical galaxy formation models. Both studies adopt a varying initial mass function and AGN energy feedback. In Fig.~\ref{overplot_wfits1}, we compare our centrally based [Z/H]--mass and [$\alpha$/Fe]--mass relations to their models.

The models by \cite{arrigoni09} are for a Chabrier IMF with slope of $x =1.1$ and a given fraction $A=0.015$ of binaries responsible for Type Ia SNe. 
The slope of the predicted mass-metallicity relation is slightly steeper than our data.
The effect of a shallower IMF is an increase in metallicity, since more massive stars are produced, thereby increasing the gas enrichment efficiency. 
There is good agreement between observed and model abundance ratios, although at higher galaxy mass the predicted values are too low. 
The lower ratio of Type Ia to Type II supernovae produces an increase in the degree of $\alpha$-elements pollution, but not in the total metallicity.
Furthermore, AGN quenching of star formation via radio mode energy feedback contributes in shortening the star formation episode of more massive galaxies.

The agreement between observed and predicted metallicities and abundance ratios by \cite{calura09b} is good, although high-mass galaxies are less metal-enriched and [$\alpha$/Fe]-enhanced than our data.
These results are obtained by implementing a star formation-dependent IMF which is a standard Salpeter with slope $x=1.35$ in objects with SFRs~$< 100~\rm M_{\odot}/\rm yr$, and with a slope of $x=1$ in objects with SFRs~$>100~\rm M_{\odot}/\rm yr$. This assumption is perhaps the main difference between the two sets of models. In particular, \cite{calura09b} stress that a constant IMF, flatter than the Salpeter, similar to that one used by \cite{arrigoni09}, has an effect on the zero point but not in the slope of the [$\alpha$/Fe]-mass relation. A star formation-dependent IMF is consistent with recent observational results of \cite{dabringhausen09}. Finally, the implementation of effects due to merger-induced starbursts and AGN energy feedback in massive galaxies is found to be essential ingredients in these models.

\section{Summary and Conclusions}
We have presented a study of stellar population properties at large galactocentric radii of early-type galaxies.
We used high-quality Gemini GMOS long-slit spectroscopic data to probe radii out to $\sim 1 - 3 r_{e}$ in 14 low-mass galaxies in the Virgo and Fornax clusters.
We extended the data sample to higher mass galaxies by introducing a novel literature compilation of stellar population properties out to $1 r_{e}$ for a sample of 37 early-type galaxies.
Stellar population radial profiles of age, total metallicity [Z/H], and [$\alpha$/Fe] abundance ratio were derived by using almost all of the Lick/IDS indices.
The comprehensive mass range uniformly covered by the final data sample ($1.6 < \log(\sigma_{0}) < 2.6$; $-16.5 > M_{B} > -23.5$; $10^{9} < \rm M_{\rm dyn}/ \rm M_{\odot} < 10^{11}$) allows us to have a clear and consistent picture of stellar population properties at large radii from dwarf to brightest cluster galaxies.

Among the stellar population parameters, the total metallicity content of a galaxy shows the strongest variation as a function of galactocentric radius.
The metallicity gradients are found to correlate with the galaxy mass. 
The mass-metallicity gradient relation shows a very different behaviour at low and high masses, with a sharp transition being seen at a dynamical mass of $\sim~3.5 \times 10^{10}$~M$_{\odot}$.

Low-mass galaxies (i.e., $1.6 < \log(\sigma_{0}) \leq 2.15$) form a tight trend with mass such that their metallicity gradients become shallower with decreasing mass.
The results imply a varying star formation efficiency with mass.
The extended old stellar discs and the prevalence of discy-shaped isophotes indicate gas dissipation and redistribution of angular momentum during an early star-forming collapse (Paper I). 
We conclude that our data are more consistent with low-mass galaxies formed in a mass-dependent early star-forming collapse. 

For galaxies with mass greater than the transition mass (i.e., $2.15 < \log(\sigma_{0}) < 2.6$), the metallicity gradients become shallower as mass increases, thus deviating from the tight low-mass trend.
We note that this transition mass value often occurs in the literature to mark a transition between physical mechanisms acting in early-type galaxies formation.
We suggest that the downturn is a consequence of the increased frequency of mergers in high-mass galaxies. More massive galaxies are expected to undergo a greater number of mergers to assemble their mass. Thus, their final gradients progressively flatten out. The different nature of merger properties (i.e, progenitors' mass ratio, the amount of gas content and the geometry of the collision) are responsible for producing the observed scatter. 

Mean stellar ages are in general slightly older than central ages.
Galaxy ages form a weak mass trend, such that more massive galaxies have older ages than low-mass systems.
We find that central and mean stellar metallicities define a clear mass-metallicty relation for our sample galaxies.
The [Z/H]$_{0}$--$\log(\sigma_{0})$ and $\langle \rm [Z/H]\rangle$--$\log(\sigma_{0})$ relationships differ not only in zero point but also in slope, such that the former trend is steeper.
The mass-metallicity gradient relation is responsible for this difference.
Massive galaxies have supersolar abundance ratios, while low-mass objects have values similar to those in stars in the solar neighbourhood.
The [$\alpha$/Fe] ratios suggest a rapid (i.e., $\leq 1$ Gyr) and efficient star formation for the high-mass galaxies.
Low-mass galaxies experience an extended (i.e., $\geq 1$ Gyr) and less efficient star-forming episode.
The negligible [$\alpha$/Fe] gradients at any given mass and out to large radii could imply similar star formation timescales for regions at different galactocentric radii.

We find that the [Z/H]--mass and [$\alpha$/Fe]--mass relationships can be explain by a mass-dependenet star formation efficiency.
This interpretation is a natural outcome of the early star-forming collapse.
In models of the hierarchical merging scenario the problem of simultaneously reproducing both the observed relationships might be solved if varying IMF, energy feedback via AGN in massive galaxies, and the effects due to merger-induced starbursts are also implemented (\citealt{arrigoni09}; \citealt{calura09}).

\section{acknowledgments}
Based on observations obtained at the Gemini Observatory, which is operated by the
Association of Universities for Research in Astronomy, Inc., under a cooperative agreement
with the NSF on behalf of the Gemini partnership: the National Science Foundation (United
States), the Science and Technology Facilities Council (United Kingdom), the
National Research Council (Canada), CONICYT (Chile), the Australian Research Council
(Australia), Minist\'{e}rio da Ci\^{e}ncia e Tecnologia (Brazil) 
and Ministerio de Ciencia, Tecnolog\'{i}a e Innovaci\'{o}n Productiva  (Argentina).
We made used of IRAF, that is distributed by the National Optical Astronomy Observatories, which are operated by the Association of Universities for Research in Astronomy, Inc., under cooperative agreement with the National Science Foundation.

MS thanks the Centre for Astrophysics \& Supercomputing for providing travel support.
We thank A. Brito, S. Brough, S. Burke-Spolaor, C. Foster, and T. Mendel for a careful reading of the manuscript.
We thank Henry Lee and the other staff astronomers at the Gemini South Telescope for the support given to our observing programs (GS-2006B-Q-74; GS-2008A-Q-3). We also thank the anonymous referee for his/her constructive comments.

\begin{appendix}

\section{Complementary data samples}
\begin{table*}
\begin{center}
\begin{tabular}{ccccccccccc}
\hline 
\hline 
Galaxy & Hubble & $r_{e}$ & P.A.& $\sigma_{0}$ & RV & Distance & $M_{B}$ & M$_{\rm dyn}$ & Ref. & Note \\
		 &  Type   & [arcsec] & [degree]&  [km s$^{-1}$] & [km s$^{-1}$] & [Mpc] &[mag] &[10$^{9}$~M$_{\odot}$] & & \\
 (1)       &  (2)            & (3) & (4) & (5) & (6) & (7) & (8) & (9) & (10) & (11) \\		 
\hline
NGC~682 & E-S0 & 32.0 & 96.0 & 199(5) &5456.2(18.3) & 73.0 & $-$19.9 & 26.9 & 4 & Isolated \\
NGC~1045 & E-S0 & 16.7 & 58.0 & 263(6) & 4496.1(12.2) & 60.0 & $-$20.9 &39.8& 4 & Isolated \\
NGC~1162 & E & 26.0 & 51.0 & 195(5) & 3789.5(11.6) & 51.0 &  $-$20.7 & 19.9 & 4 & Isolated \\
NGC~1400 & E1 & 27.0 & 40.0 &281(7) & 560.1(10.7) & 26.42 & $-$20.2 & 33.8 & 1 & BGG\\
NGC~1407 & E0	 & 72.0 & 35.0 &270(7) & 1794.1(10.9) & 28.84 & $-$21.6 & 91.2 & 1 & BGG\\
NGC~1600 & E3 & 47.5 & 5.2 & 336(8) & 4708.4(20.0) & 62.23  & $-$22.0 & 194.9 & 2 & Field\\
NGC~1700 & E4	 & 13.7 & 87.0 & 235(4) & 3891.2(17.2) & 51.52 & $-$21.4	 & 23.4 & 2 & Group\\
NGC~2271 & E-S0 & 19.0 & 84.0 & 234(5) & 2402.3(9.0) & 32.0 & $-$20.0 &19.8 & 4 & Isolated \\
NGC~2831 & E & 23.0 & 166.3 & 202(4) & 5149.5(13.0) & 68.0 &$-$19.9 & 39.8 & 5 & Group \\
NGC~2832 & E & 25.0 & 171.8 & 288(5) & 6871.6(16.0) & 91.0 & $-$22.1 & 114.8 & 5 & Group \\
NGC~2865 & E & 27.7 & 153.0 & 178(8) & 2472.5(10.4) & 35.0 & $-$20.7 & 18.6 & 4 & Isolated \\
NGC~3226 & E & 34.0 & 12.0 & 203(4) & 1304.7(10) & 23.4 & $-$19.9 & 22.9 & 5 & Group \\
NGC~3377 & E5-6 & 33.7 & 37.2 & 139(3) & 684.2(11.3) & 10.28 & $-$18.8 & 3.9 & 2 & Group\\
NGC~3379 & E1 & 35.2 & 71.0 & 207(2) & 897.1(5.8) & 10.28 & $-$19.8 & 9.5 & 2 & Group\\
NGC~3384 & E-S0 & 24.9 & 53.0 & 148(3) & 903.1(10.2) & 10.28 & $-$19.2 & 3.4 & 2 & Group\\
NGC~3557 & E & 36.7 & 31.0 & 282(16) & 3072.1(8.0) & 44.2 & $-$21.8 & 77.6 & 3 & BGG\\
NGC~3608 & E & 34.0 & 80.0 & 208(3) & 1198.3(24.3) & 23.4 & $-$20.5 & 23.9 & 5 & Group\\
NGC~3640 & E & 36.6 & 80.0 & 154(7) & 1341.4(5.4) & 19.2 &$-$20.0 & 14.1 & 3 & BGG\\
NGC~4240 & E & 27.2 & 276.0 & 123(3) & 1916.7(12.4) & 26.0 & $-$18.8 & 3.6 & 4 & Isolated \\
NGC~4291 & E & 17.0 & 107.7 & 292(4) & 1731.7(3.0) & 29.4 & $-$19.9 & 24.5 & 5 & Cluster\\
NGC~4365 & E & 50.0 & 40.3 &  254(3) & 1240.2(7.3) & 16.8 & $-$20.5 & 30.1 & 5 & Cluster\\
NGC~4374 & E & 51.0 & 132.9 & 316(5) & 1014.4(4.4) & 16.8 & $-$20.9 & 46.7 & 5 & Cluster\\
NGC~4472 & E2/S0 & 104.0 & 157.1 & 291.1(3) & 974(2.9) & 17.13 & $-$21.8 & 89.1 & 2 & Cluster\\
NGC~4552 & E & 29.0 & 64.0 & 272(4) & 357.2(10.6) & 16.8 & $-$20.3 & 19.9 & 5 & Cluster\\
NGC~4636 & E & 89.0 & 149.6 & 243(3) & 933.4(8.3) & 17.0 & $-$20.6 & 48.9 & 5 & Cluster\\
NGC~4697 & E & 67.0 & 83.1 & 194(2) & 1240.3(3.3) & 23.3 & 	$-$20.8 & 22.9 & 5 & Cluster\\
NGC~5044 & E & 24.3 & 20.0 & 223(6) & 2755.6(12.2) & 39.6 & $-$21.5 &27.5 & 3 & BGG\\
NGC~5322 & E & 34.0 & 95.0 & 233(3) &1781.2(43.3) & 31.6 & $-$21.2 & 31.6 & 5 & Cluster\\
Abell~754 & E & 10.0 & 96.0 & 294(9) & 16468.2(11.3) & 245.1 & $-$22.6 &	123.0& 3 & BCG\\
Abell~970 & E & 4.4 & 177.0 & 272(14) & 17512.2(6.2) & 261.3 & $-$20.9 & 50.1& 3 & BCG\\
Abell~978 & E & 8.4 & 18.0 & 309(13) & 17695.1(15.1) & 264.1 & $-$21.9 & 123.0 & 3 & BCG\\
ESO~153-G003 & E & 10.2 & 291 & 219(5) & 6242.6(7.1) & 84.0 & $-$20.9 & 72.4 & 4 & Isolated\\
ESO~218-G002 & E & 20.1 & 3.0 & 269(6) & 4072.7(11.1) & 54.0 &	$-$20.9 & 26.9 & 4 & Isolated\\
ESO~318-G021 & E & 21.7 & 121.0 & 234(5) & 4641.2(12.2) &62.0 &	$-$20.7 & 39.8 & 4 & Isolated\\
MCG-01-27-013 & E & 13.8 & 11.0 & 246(6) & 9044.2(9.0) & 121.0 & $-$21.4 & 72.4 &4 & Isolated\\
MCG-02-13-009 & E & 23.2 & 149.0 & 214(10) & 5494.0(8.0) & 73.0 & $-$21.3 & 26.9 & 4 & Isolated\\
MCG-03-26-030 & E & 15.6 & 104.0 & 295(7) & 8949.0(7.0) & 119.0 & $-$21.7 & 91.2 &4 & Isolated\\
\hline \hline
\end{tabular}
\end{center}
\caption[]{Galaxy properties of the comparison data sample. Col(1): name of the galaxy; (2): morphological type from the HyperLEDA database; (3): B-band effective radius along the galaxy semi-major axis; (4): position angle of the major axis; (5, 6): central stellar velocity dispersion and radial velocity; (7): distance of the galaxy; (8): total $B$-band absolute magnitudes estimated from total apparent magnitudes given in the NASA/IPAC Extragalactic Database and corrected for Galactic extinction; (9): dynamical mass of the galaxy using the $\log$(M$_{\rm dyn}$)~=~2$\log(\sigma_{0}) + \log(r_{e})+ 5.0$, where $r_{e}$ is the effective radius expressed in parsecs (\citealt{cappellari06}); (10): reference work; Ref(1): \cite{spolaor08a}; Ref(2): \cite{patrizia07}; Ref(3): \cite{brough07}; Ref(4):\cite{reda07}; Ref(5): \cite{proctor03}; col(11): Note on individual galaxies}
\label{gal2_prop}
\end{table*}

\section{Individual galaxies}
\label{individual}

\subsection{FCC~148 (NGC 1375)}
The stellar population parameters radial profiles extend out to two effective radii.
We measured an age gradient of $0.22~\pm~0.05$~dex~per~decade and a central age of Age$_{0}=~1.7$~Gyr.
The galaxy is observed to be metal rich in the central region and to have a shallow negative radial gradient, with a mean metallicity of $\langle \rm [Z/H] \rangle=-0.11$.
The [$\alpha$/Fe] radial profile is flat and we found a mean $\alpha$$-$abundance ratio of $-0.14$.
\subsection{FCC~153 (IC 1963)}
The radial profiles of the stellar population parameters reach almost four effective radii. 
The age and $\alpha$$-$abundance ratio profiles are flat and they have mean values of $\langle \rm Age \rangle=3.8$~Gyr and $\langle \rm [\alpha/Fe] \rangle=0.04$.
Interestingly, we found that the total metallicity content increases with increasing galactocentric radius, from a starting central value of $[Z/H]_{0}=~0.29$.
We measured a positive metallicity gradient of $0.07~\pm~0.04$~dex~per~decade. 
\subsection{FCC~170 (NGC 1381)}
We studied the stellar population parameters out to almost four effective radii. 
The age profile is uniformly old, with $\langle \rm Age \rangle=11.8$~Gyr.
The [Z/H] profile has a steep negative gradient out to the largest radius, grad~[Z/H] = $-0.34~\pm~0.02$~dex~per~decade.
The enrichment by $\alpha$$-$elements is solar, $\langle \rm [\alpha/Fe] \rangle=0.02$, and does not present a statistically significant radial gradient.
\subsection{FCC~277 (NGC 1428)}
The radial extent of our stellar population analysis reaches two effective radii.
The age profile is uniformly old at all radii.
The galaxy is observed to be metal poor, $\langle \rm [Z/H] \rangle=-0.30$ with a shallow negative metallicity gradient of $-0.16~\pm~0.02$~dex~per~decade.
The amount of enrichment by $\alpha$$-$elements increases with increasing radius, due to a shallow positive radial gradients of $0.10~\pm~0.04$~dex~per~decade.
\subsection{FCC~301(ESO 358-G059)}
The stellar population analysis extends out to one effective radius.
The galaxy is observed to have a constant age profile with a mean age of $\langle \rm Age \rangle=5.0$~Gyr.
The [Z/H] radial profile is observed to be flat with a mean value of $\langle \rm [Z/H] \rangle=-0.05$.
The [$\alpha$/Fe] radial profile is constant with a mean value of 0.04.
\subsection{FCC~335 (ESO 359-G002)}
The stellar population parameters have been observed out to one effective radius.
The galaxy has a significant age gradient of $0.47~\pm~0.07$~dex~per~decade, with a central age of Age$_{0}=~1.5$~Gyr.
We measured a positive metallicity gradient of $0.02~\pm~0.04$~dex~per~dex. 
We found that the galaxy is metal poor with central and mean values of [Z/H]$_{0}=~-0.50$ and $\langle \rm [Z/H] \rangle=-0.42$.
The galaxy has a solar abundance of $\alpha$$-$elements, $\rm [\alpha/Fe]_{0}=~-0.04$, and a positive radial gradient.
\subsection{VCC~575 (NGC 4318)}
The stellar population parameters radial profiles extend to almost two effective radii. 
The lack of age gradient implies that the galaxy is constantly old at all radii.
We measured a mean age of 11.8~Gyr.
The galaxy is observed to be metal poor, $\langle \rm [Z/H] \rangle=-0.30$, with a steep negative metallicity gradient. 
The central $\alpha$$-$abundance ratio is solar, but increase to $\langle \rm [\alpha/Fe] \rangle=0.15$ in the outer galactic regions.
\subsection{VCC~828 (NGC 4387)}
We measured stellar population radial profiles out to almost two effective radii. 
The age profile is uniformly old, with a mean age of $\langle \rm Age \rangle=13.3$~Gyr.
We found a metallicity gradient of $-0.21~\pm~0.02$~dex~per~decade, from a central metallicity of [Z/H]$_{0}=~0.09$.
The [$\alpha$/Fe] abundance content increases with increasing galactocentric radius, with a central value of $\rm [\alpha/Fe]_{0}=~0.11$.
\subsection{VCC~1025 (NGC 4434)}
We spatially resolved galactic regions out to almost two effective radii. 
We detected a positive age gradient along the galaxy's radius of $0.13~\pm~0.02$~dex~per~decade.
The central age value is Age$_{0}=~9.5$~Gyr.
The metallicity gradient of $-0.33~\pm~0.02$~dex~per~decade is the steepest observed in our sample galaxies.
The mean $\alpha$$-$abundance content is supersolar, $\langle \rm [\alpha/Fe] \rangle=0.22$ and we detected a statistically significant positive radial gradient.
\subsection{VCC~1146 (NGC 4458)}
The radial extent of our stellar population analysis reach one effective radius. 
The age radial profile is observed to be uniform with a mean age of 12.0~Gyr.
The [Z/H] and [$\alpha$/Fe] radial profiles are observed to decrease with increasing galactocentric radius, indicative of negative radial gradients: grad~[Z/H] = $-0.22~\pm~0.02$~dex~per~dex and grad~$ \rm [\alpha/Fe] = -0.07 \pm~0.02$~dex~per~decade.
The galaxy is observed to be metal poor and significantly polluted by $\alpha$$-$elements. 
We measured $\rm [Z/H]_{0}=~-0.17$ and $\rm [\alpha/Fe]_{0}=~0.25$.
\subsection{VCC~1178 (NGC 4464)}
The radial profiles of the stellar population parameters are studied out to three effective radii.
The galaxy is constantly old along the galactocentric radius, with a mean age of $\langle Age \rangle=12.8$~Gyr.
We measured statistically significant [Z/H] and [$\alpha$/Fe] negative radial gradients: grad~$\rm [Z/H] = -0.22~\pm~0.025$~dex~per~decade and grad~$\rm [\alpha/Fe] = -0.07~\pm~0.02$~dex~per~decade.
The galaxy is found to be metal poor, $\langle \rm [Z/H] \rangle = -0.22$ and chemically enriched by a supersolar amount of $\alpha$$-$elements, $\langle \rm [\alpha/Fe] \rangle=0.12$.
\subsection{VCC~1297 (NGC 4486B)}
 We have been able to spatially resolved galactic regions out to four effective radii.
The age radial profile is uniform at all radii and we measured $\langle \rm Age \rangle=13.0$~Gyr.
Total metallicity and $\alpha$$-$abundance ratio content decrease with increasing galactocentric radius: grad~$\rm [Z/H] = -0.23~\pm~0.02$~dex~per~decade and grad~$\rm [\alpha/Fe] = -0.06~\pm~0.02$~dex~per~decade.
Central and mean metallicity are significantly different at $\rm [Z/H]_{0} = 0.31$ and $\langle \rm [Z/H] \rangle=-0.01$.
\subsection{VCC~1475 (NGC 4515)}
The stellar population analysis extends to two effective radii. 
The age radial profile has a statistically significant positive gradient of $0.04~\pm~0.02$~dex~per~decade.
The mean age of the galaxy is of $\langle \rm Age \rangle=12.0$~Gyr.
The galaxy has a negative radial gradient and is metal poor, $\rm [Z/H]_{0} = -0.14$.
The [$\alpha$/Fe] radial profile is constant along the galactocentric radius with a mean value of $\langle \rm [\alpha/Fe] \rangle=0.12$.
\subsection{VCC~1630 (NGC 4551)}
The stellar population parameters radial profiles are studied out to two effective radii.
The age profile is observed to slightly increase with increasing radius, from a central age value of $\rm Age_{0} = 10.5$~Gyr.
The galaxy is observed to be metal rich in the central regions, $\rm [Z/H]_{0}=~0.22$.
The metallicity radial profile has a negative gradient of $-0.25~\pm~0.02$~dex~per~decade.
The [$\alpha$/Fe] radial profile is uniform along the galactocentric radius with a mean value of $\langle \rm [\alpha/Fe] \rangle=0.16$.

\section{Stellar population parameters}
\begin{table*}
\begin{center}
\begin{tabular}{ccccc}
\hline 
\hline 
Galaxy & $\log(\sigma_{0})$ & grad~Age &  grad~[Z/H] & grad~[$\alpha$/Fe] \\
  (1)       &  (2)            & (3) & (4) & (5) \\
\hline 
FCC~148&  1.84 $\pm$ 0.02 &  0.22 $\pm$ 0.05 & -0.06 $\pm$ 0.04 &  0.05 $\pm$ 0.04  \\      
FCC~153&  1.68 $\pm$ 0.02 &  0.02 $\pm$ 0.06 &  0.07 $\pm$ 0.04 &  0.01 $\pm$ 0.03  \\ 	
FCC~170&  2.18 $\pm$ 0.01 &  0.05 $\pm$ 0.04 & -0.34 $\pm$ 0.02 &  0.05 $\pm$ 0.05  \\ 	 
FCC~277&  1.97 $\pm$ 0.01 &  0.03 $\pm$ 0.04 & -0.16 $\pm$ 0.02 &  0.10 $\pm$ 0.04  \\ 	 
FCC~301&  1.77 $\pm$ 0.02 & -0.02 $\pm$ 0.05 & -0.01 $\pm$ 0.04 &  0.01 $\pm$ 0.04  \\ 	
FCC~335&  1.65 $\pm$ 0.03 &  0.47 $\pm$ 0.07 &  0.02 $\pm$ 0.04 &  0.14 $\pm$ 0.04  \\      
VCC~575 &  1.94 $\pm$ 0.01 & -0.02 $\pm$ 0.02 & -0.11 $\pm$ 0.02 &  0.08 $\pm$ 0.04  \\      
VCC~828 &  2.02 $\pm$ 0.01 &  0.08 $\pm$ 0.02 & -0.21 $\pm$ 0.02 &  0.06 $\pm$ 0.02  \\ 	 
VCC~1025 &  2.13 $\pm$ 0.01 &  0.13 $\pm$ 0.02 & -0.33 $\pm$ 0.02 &  0.06 $\pm$ 0.02  \\ 	 
VCC~1146 &  2.01 $\pm$ 0.01 &  0.03 $\pm$ 0.01 & -0.22 $\pm$ 0.02 & -0.07 $\pm$ 0.02  \\ 	 
VCC~1178 &  2.03 $\pm$ 0.01 & -0.01 $\pm$ 0.01 & -0.22 $\pm$ 0.02 & -0.07 $\pm$ 0.02  \\ 	 
VCC~1297 &  2.43 $\pm$ 0.01 &  0.03 $\pm$ 0.02 & -0.23 $\pm$ 0.02 & -0.06 $\pm$ 0.02  \\ 	 
VCC~1475 &  1.96 $\pm$ 0.01 &  0.04 $\pm$ 0.02 & -0.16 $\pm$ 0.02 & -0.03 $\pm$ 0.02  \\ 	 
VCC~1630 &  2.04 $\pm$ 0.01 &  0.06 $\pm$ 0.02 & -0.25 $\pm$ 0.02 &  0.02 $\pm$ 0.02  \\ 	 
NGC~682      &  2.43 $\pm$ 0.01 & -0.00 $\pm$ 0.02 & -0.38 $\pm$ 0.04 &  0.02 $\pm$ 0.02  \\         
NGC~1045     &  2.45 $\pm$ 0.01 & -0.02 $\pm$ 0.02 & -0.47 $\pm$ 0.04 &  0.04 $\pm$ 0.03  \\         
NGC~1162     &  2.53 $\pm$ 0.01 &  0.00 $\pm$ 0.00 & -0.19 $\pm$ 0.02 &  0.01 $\pm$ 0.05  \\        
NGC~1400     &  2.37 $\pm$ 0.01 &  0.12 $\pm$ 0.05 & -0.34 $\pm$ 0.01 & -0.01 $\pm$ 0.01  \\        
NGC~1407     &  2.14 $\pm$ 0.01 &  0.24 $\pm$ 0.05 & -0.47 $\pm$ 0.01 & -0.08 $\pm$ 0.01  \\       
NGC~1600     &  2.32 $\pm$ 0.01 &  0.00 $\pm$ 0.00 & -0.28 $\pm$ 0.01 &  0.04 $\pm$ 0.01  \\       
NGC~1700     &  2.17 $\pm$ 0.01 &  0.09 $\pm$ 0.04 & -0.38 $\pm$ 0.01 &  0.03 $\pm$ 0.01  \\       
NGC~2271     &  2.46 $\pm$ 0.01 &  0.00 $\pm$ 0.01 & -0.19 $\pm$ 0.01 &  0.06 $\pm$ 0.01  \\       
NGC~2831     &  2.45 $\pm$ 0.02 &  0.11 $\pm$ 0.07 & -0.40 $\pm$ 0.03 &  0.11 $\pm$ 0.02  \\            
NGC~2832     &  2.26 $\pm$ 0.02 &  0.07 $\pm$ 0.06 & -0.20 $\pm$ 0.04 &  0.06 $\pm$ 0.03  \\         
NGC~2865     &  2.34 $\pm$ 0.01 & -0.09 $\pm$ 0.08 & -0.12 $\pm$ 0.08 & -0.03 $\pm$ 0.06  \\           
NGC~3226     &  2.46 $\pm$ 0.01 & -0.09 $\pm$ 0.07 & -0.54 $\pm$ 0.10 &  0.05 $\pm$ 0.05  \\           
NGC~3377     &  2.43 $\pm$ 0.02 & -0.02 $\pm$ 0.03 & -0.58 $\pm$ 0.10 &  0.05 $\pm$ 0.03  \\           
NGC~3379     &  2.48 $\pm$ 0.02 &  0.06 $\pm$ 0.03 & -0.20 $\pm$ 0.04 &  0.14 $\pm$ 0.05  \\          
NGC~3384     &  2.30 $\pm$ 0.01 & -0.03 $\pm$ 0.08 & -0.25 $\pm$ 0.03 & -0.01 $\pm$ 0.03  \\     
NGC~3557     &  2.42 $\pm$ 0.01 & -0.19 $\pm$ 0.04 & -0.24 $\pm$ 0.03 & -0.05 $\pm$ 0.03  \\     
NGC~3608     &  2.29 $\pm$ 0.01 &  0.29 $\pm$ 0.07 & -0.44 $\pm$ 0.03 & -0.02 $\pm$ 0.04  \\     
NGC~3640     &  2.37 $\pm$ 0.01 &  0.03 $\pm$ 0.07 & -0.05 $\pm$ 0.04 &  0.07 $\pm$ 0.05  \\     
NGC~4240     &  2.25 $\pm$ 0.02 &  0.63 $\pm$ 0.07 & -0.47 $\pm$ 0.05 & -0.03 $\pm$ 0.03  \\     
NGC~4291     &  2.09 $\pm$ 0.01 &  0.16 $\pm$ 0.12 & -0.30 $\pm$ 0.06 & -0.01 $\pm$ 0.06  \\     
NGC~4365     &  2.34 $\pm$ 0.01 &  0.02 $\pm$ 0.08 & -0.22 $\pm$ 0.05 &  0.03 $\pm$ 0.07  \\     
NGC~4374     &  2.43 $\pm$ 0.01 & -0.04 $\pm$ 0.04 & -0.28 $\pm$ 0.04 & -0.04 $\pm$ 0.07  \\     
NGC~4472     &  2.37 $\pm$ 0.01 & -0.16 $\pm$ 0.11 & -0.01 $\pm$ 0.07 &  0.06 $\pm$ 0.07  \\     
NGC~4552     &  2.39 $\pm$ 0.01 &  0.07 $\pm$ 0.25 & -0.05 $\pm$ 0.14 & -0.14 $\pm$ 0.13  \\     
NGC~4636     &  2.33 $\pm$ 0.02 & -0.58 $\pm$ 0.27 & -0.15 $\pm$ 0.15 & -0.13 $\pm$ 0.10  \\     
NGC~4697     &  2.47 $\pm$ 0.01 &  0.01 $\pm$ 0.04 & -0.10 $\pm$ 0.02 &  0.05 $\pm$ 0.08  \\     
NGC~5044     &  2.30 $\pm$ 0.01 &  0.24 $\pm$ 0.23 & -0.26 $\pm$ 0.12 & -0.01 $\pm$ 0.09  \\             
NGC~5322     &  2.45 $\pm$ 0.01 &  0.08 $\pm$ 0.01 & -0.20 $\pm$ 0.04 &  0.00 $\pm$ 0.05  \\    	
Abell~754    &  2.30 $\pm$ 0.01 & -0.14 $\pm$ 0.02 & -0.27 $\pm$ 0.06 & -0.12 $\pm$ 0.06  \\       
Abell~970    &  2.31 $\pm$ 0.01 & -0.16 $\pm$ 0.11 & -0.15 $\pm$ 0.07 & -0.07 $\pm$ 0.02  \\       
Abell~978    &  2.46 $\pm$ 0.01 & -0.11 $\pm$ 0.08 & -0.17 $\pm$ 0.05 & -0.08 $\pm$ 0.02  \\       
ESO~153-G003 &  2.40 $\pm$ 0.01 & -0.06 $\pm$ 0.07 & -0.13 $\pm$ 0.04 & -0.00 $\pm$ 0.02  \\       
ESO~218-G002 &  2.49 $\pm$ 0.01 & -0.08 $\pm$ 0.02 & -0.14 $\pm$ 0.02 &  0.00 $\pm$ 0.04  \\       
ESO~318-G021 &  2.43 $\pm$ 0.01 &  0.10 $\pm$ 0.03 & -0.21 $\pm$ 0.06 &  0.07 $\pm$ 0.03  \\        
MCG-01-27-013&  2.38 $\pm$ 0.01 &  0.10 $\pm$ 0.06 & -0.22 $\pm$ 0.05 & -0.01 $\pm$ 0.01  \\        
MCG-02-13-009&  2.28 $\pm$ 0.01 &  0.07 $\pm$ 0.06 & -0.29 $\pm$ 0.04 & -0.03 $\pm$ 0.05  \\        
MCG-03-26-030&  2.36 $\pm$ 0.01 &  0.12 $\pm$ 0.01 & -0.25 $\pm$ 0.03 &  0.04 $\pm$ 0.06  \\        
\hline 
\hline
\end{tabular}
\end{center}
\caption[]{Stellar population gradients for our sample galaxies.}
\label{Pop_par1}
\end{table*} 

\begin{table*}
\begin{center}
\begin{tabular}{ccccc}
\hline 
\hline 
Galaxy & $\log(\sigma_{0})$ & $\log$(Age$_{0}$) & [Z/H]$_{0}$  & [$\alpha$/Fe]$_{0}$ \\
  (1)       &  (2)            & (3) & (4) & (5) \\
\hline 
FCC~148&   1.84 $\pm$ 0.02 & 0.23 $\pm$ 0.05 &  0.00 $\pm$ 0.06 &   -0.15 $\pm$ 0.05     \\
FCC~153&   1.68 $\pm$ 0.02 & 0.54 $\pm$ 0.03 &  0.29 $\pm$ 0.03 &    0.03 $\pm$ 0.02     \\
FCC~170&   2.18 $\pm$ 0.01 & 1.02 $\pm$ 0.03 &  0.15 $\pm$ 0.03 &    0.00 $\pm$ 0.02     \\
FCC~277&   1.97 $\pm$ 0.01 & 1.00 $\pm$ 0.03 & -0.01 $\pm$ 0.04 &    0.01 $\pm$ 0.02     \\
FCC~301&   1.77 $\pm$ 0.02 & 0.68 $\pm$ 0.04 & -0.05 $\pm$ 0.03 &   -0.02 $\pm$ 0.04     \\
FCC~335&   1.65 $\pm$ 0.03 & 0.16 $\pm$ 0.18 & -0.49 $\pm$ 0.08 &   -0.04 $\pm$ 0.12     \\
VCC~575 &   1.94 $\pm$ 0.01 & 1.03 $\pm$ 0.05 & -0.08 $\pm$ 0.05 &    0.02 $\pm$ 0.03     \\
VCC~828 &   2.02 $\pm$ 0.01 & 1.06 $\pm$ 0.05 &  0.09 $\pm$ 0.05 &    0.11 $\pm$ 0.03     \\
VCC~1025 &   2.13 $\pm$ 0.01 & 0.98 $\pm$ 0.06 &  0.15 $\pm$ 0.06 &    0.15 $\pm$ 0.03     \\
VCC~1146 &   2.01 $\pm$ 0.01 & 1.13 $\pm$ 0.04 & -0.17 $\pm$ 0.07 &    0.25 $\pm$ 0.03     \\
VCC~1178 &   2.03 $\pm$ 0.01 & 1.13 $\pm$ 0.02 & -0.02 $\pm$ 0.02 &    0.24 $\pm$ 0.02     \\
VCC~1297 &   2.43 $\pm$ 0.01 & 1.10 $\pm$ 0.04 &  0.30 $\pm$ 0.05 &    0.35 $\pm$ 0.04     \\
VCC~1475 &   1.96 $\pm$ 0.01 & 1.06 $\pm$ 0.04 & -0.13 $\pm$ 0.03 &    0.18 $\pm$ 0.04     \\
VCC~1630 &   2.04 $\pm$ 0.01 & 1.02 $\pm$ 0.04 &  0.21 $\pm$ 0.04 &    0.14 $\pm$ 0.02     \\
NGC~682      &   2.43 $\pm$ 0.01 & 1.08 $\pm$ 0.04 &  0.29 $\pm$ 0.08 &    0.41 $\pm$ 0.05     \\
NGC~1045     &   2.45 $\pm$ 0.01 & 1.14 $\pm$ 0.03 &  0.25 $\pm$ 0.06 &    0.33 $\pm$ 0.04     \\
NGC~1162     &   2.53 $\pm$ 0.01 & 1.13 $\pm$ 0.01 &  0.32 $\pm$ 0.01 &    0.36 $\pm$ 0.01    \\
NGC~1400     &   2.37 $\pm$ 0.01 & 0.46 $\pm$ 0.01 &  0.62 $\pm$ 0.01 & 	0.18 $\pm$ 0.01   \\
NGC~1407     &   2.14 $\pm$ 0.01 & 0.74 $\pm$ 0.01 &  0.34 $\pm$ 0.01 &     0.27 $\pm$ 0.01    \\
NGC~1600     &   2.32 $\pm$ 0.01 & 1.15 $\pm$ 0.01 &  0.24 $\pm$ 0.01 &     0.28 $\pm$ 0.01    \\
NGC~1700     &   2.17 $\pm$ 0.01 & 0.48 $\pm$ 0.01 &  0.59 $\pm$ 0.01 & 	0.18 $\pm$ 0.01    \\
NGC~2271     &   2.46 $\pm$ 0.01 & 1.05 $\pm$ 0.01 &  0.35 $\pm$ 0.01 & 	0.28 $\pm$  0.01   \\
NGC~2831     &   2.45 $\pm$ 0.02 & 1.05 $\pm$ 0.03 &  0.25 $\pm$ 0.01 & 	0.21 $\pm$  0.02   \\
NGC~2832     &   2.26 $\pm$ 0.02 & 0.83 $\pm$ 0.03 &  0.15 $\pm$ 0.02 & 	0.21 $\pm$  0.02   \\
NGC~2865     &   2.34 $\pm$ 0.01 & 1.18 $\pm$ 0.02 &  0.10 $\pm$ 0.02 & 	0.36 $\pm$  0.03   \\
NGC~3226     &   2.46 $\pm$ 0.01 & 1.18 $\pm$ 0.02 &  0.20 $\pm$ 0.01 & 	0.24 $\pm$  0.03   \\
NGC~3377     &   2.43 $\pm$ 0.02 & 1.18 $\pm$ 0.02 &  0.25 $\pm$ 0.02 & 	0.27 $\pm$  0.03   \\
NGC~3379     &   2.48 $\pm$ 0.02 & 1.12 $\pm$ 0.03 &  0.05 $\pm$ 0.04 & 	0.06 $\pm$  0.04   \\
NGC~3384     &   2.30 $\pm$ 0.01 & 0.90 $\pm$ 0.03 &  0.33 $\pm$ 0.02 &     0.30 $\pm$  0.01   \\
NGC~3557     &   2.42 $\pm$ 0.01 & 1.02 $\pm$ 0.02 &  0.31 $\pm$ 0.02 & 	0.36 $\pm$  0.02   \\
NGC~3608     &   2.29 $\pm$ 0.01 & 0.83 $\pm$ 0.02 &  0.29 $\pm$ 0.02 & 	0.31 $\pm$  0.01   \\
NGC~3640     &   2.37 $\pm$ 0.01 & 1.06 $\pm$ 0.02 &  0.44 $\pm$ 0.02 & 	0.14 $\pm$  0.01   \\
NGC~4240     &   2.25 $\pm$ 0.02 & 0.23 $\pm$ 0.02 &  0.48 $\pm$ 0.05 & 	0.07 $\pm$  0.01   \\
NGC~4291     &   2.09 $\pm$ 0.01 & 0.87 $\pm$ 0.02 &  0.23 $\pm$ 0.02 &   -0.07 $\pm$  0.01    \\
NGC~4365     &   2.34 $\pm$ 0.01 & 1.05 $\pm$ 0.02 &  0.19 $\pm$ 0.01 & 	0.28 $\pm$  0.01   \\
NGC~4374     &   2.43 $\pm$ 0.01 & 1.17 $\pm$ 0.02 &  0.35 $\pm$ 0.01 & 	0.26 $\pm$  0.01   \\
NGC~4472     &   2.37 $\pm$ 0.01 & 0.95 $\pm$ 0.03 &  0.34 $\pm$ 0.03 & 	0.08 $\pm$  0.01   \\
NGC~4552     &   2.39 $\pm$ 0.01 & 0.90 $\pm$ 0.02 &  0.34 $\pm$ 0.04 &   -0.03 $\pm$  0.01    \\
NGC~4636     &   2.33 $\pm$ 0.02 & 0.98 $\pm$ 0.02 &  0.36 $\pm$ 0.02 & 	0.05 $\pm$  0.02   \\
NGC~4697     &   2.47 $\pm$ 0.01 & 1.16 $\pm$ 0.03 &  0.27 $\pm$ 0.01 & 	0.15 $\pm$  0.03   \\
NGC~5044     &   2.30 $\pm$ 0.01 & 0.61 $\pm$ 0.06 &  0.31 $\pm$ 0.08 &  	0.25 $\pm$  0.03   \\
NGC~5322     &   2.45 $\pm$ 0.01 & 0.87 $\pm$ 0.06 &  0.45 $\pm$ 0.07 &  	0.30 $\pm$  0.02   \\
Abell~754    &   2.30 $\pm$ 0.01 & 1.02 $\pm$ 0.05 &  0.32 $\pm$ 0.06 &  	0.42 $\pm$  0.02   \\
Abell~970    &   2.31 $\pm$ 0.01 & 0.95 $\pm$ 0.04 &  0.38 $\pm$ 0.03 &  	0.27 $\pm$  0.01   \\
Abell~978    &   2.46 $\pm$ 0.01 & 1.01 $\pm$ 0.05 &  0.28 $\pm$ 0.06 &  	0.30 $\pm$  0.02   \\
ESO~153-G003 &   2.40 $\pm$ 0.01 & 0.98 $\pm$ 0.04 &  0.41 $\pm$ 0.03 &  	0.25 $\pm$  0.01   \\
ESO~218-G002 &   2.49 $\pm$ 0.01 & 1.13 $\pm$ 0.05 &  0.28 $\pm$ 0.06 &  	0.30 $\pm$  0.02   \\
ESO~318-G021 &   2.43 $\pm$ 0.01 & 0.98 $\pm$ 0.04 &  0.45 $\pm$ 0.06 &  	0.32 $\pm$  0.02   \\
MCG-01-27-013&   2.38 $\pm$ 0.01 & 0.91 $\pm$ 0.06 &  0.43 $\pm$ 0.06 &  	0.27 $\pm$  0.02   \\
MCG-02-13-009&   2.28 $\pm$ 0.01 & 0.71 $\pm$ 0.07 &  0.48 $\pm$ 0.04 &  	0.20 $\pm$  0.02   \\
MCG-03-26-030&   2.36 $\pm$ 0.01 & 0.62 $\pm$ 0.05 &  0.39 $\pm$ 0.04 &  	0.15 $\pm$  0.02   \\
\hline 
\hline
\end{tabular}
\end{center}
\caption[]{Central values of stellar population parameters for our sample galaxies.}
\label{Pop_par2}
\end{table*} 

\begin{table*}
\begin{center}
\begin{tabular}{ccccc}
\hline 
\hline 
Galaxy & $\log(\sigma_{0})$ & $\langle \log(\rm Age) \rangle$ & $\langle \rm [Z/H] \rangle$& $\langle \rm [\alpha/Fe] \rangle$\\
  (1)       &  (2)            & (3) & (4) & (5) \\
\hline 
VCC~575      &  1.94 $\pm$ 0.01 &   1.07 $\pm$ 0.09 &   -0.30 $\pm$ 0.09 &    0.15 $\pm$ 0.09 \\
VCC~828      &  2.02 $\pm$ 0.01 &   1.12 $\pm$ 0.09 &   -0.26 $\pm$ 0.09 &    0.24 $\pm$ 0.09 \\  
VCC~1025      &  2.13 $\pm$ 0.01 &   1.11 $\pm$ 0.09 &   -0.31 $\pm$ 0.09 &    0.22 $\pm$ 0.09 \\ 
VCC~1146    &  2.01 $\pm$ 0.01 &   1.08 $\pm$ 0.09 &   -0.51 $\pm$ 0.09 &    0.10 $\pm$ 0.09 \\ 
VCC~1178      &  2.03 $\pm$ 0.01 &   1.11 $\pm$ 0.09 &   -0.22 $\pm$ 0.09 &    0.12 $\pm$ 0.09 \\ 
VCC~1297      &  2.43 $\pm$ 0.01 &   1.15 $\pm$ 0.09 &    0.01 $\pm$ 0.09 &    0.25 $\pm$ 0.09 \\ 
VCC~1475      &  1.96 $\pm$ 0.01 &   1.07 $\pm$ 0.09 &   -0.45 $\pm$ 0.09 &    0.12 $\pm$ 0.09 \\ 
VCC~1630      &  2.04 $\pm$ 0.01 &   1.10 $\pm$ 0.09 &   -0.10 $\pm$ 0.09 &    0.16 $\pm$ 0.09 \\ 
FCC~148     &  1.84 $\pm$ 0.02 &   0.71 $\pm$ 0.09 &   -0.27 $\pm$ 0.09 &   -0.13 $\pm$ 0.09 \\
FCC~153     &  1.68 $\pm$ 0.02 &   0.57 $\pm$ 0.09 &    0.37 $\pm$ 0.09 &    0.04 $\pm$ 0.09 \\
FCC~170     &  2.18 $\pm$ 0.01 &   1.07 $\pm$ 0.09 &   -0.12 $\pm$ 0.09 &    0.01 $\pm$ 0.09 \\  
FCC~277     &  1.97 $\pm$ 0.01 &   1.05 $\pm$ 0.09 &   -0.30 $\pm$ 0.09 &    0.18 $\pm$ 0.09 \\  
FCC~301     &  1.77 $\pm$ 0.02 &   0.70 $\pm$ 0.09 &   -0.05 $\pm$ 0.09 &    0.04 $\pm$ 0.09 \\
FCC~335     &  1.65 $\pm$ 0.03 &   0.73 $\pm$ 0.09 &   -0.42 $\pm$ 0.09 &    0.15 $\pm$ 0.09 \\
NGC~682      &  2.43 $\pm$ 0.01 &   1.07 $\pm$ 0.09 &   -0.20 $\pm$ 0.09 &    0.42 $\pm$ 0.09 \\
NGC~1045     &  2.45 $\pm$ 0.01 &   1.12 $\pm$ 0.09 &   -0.50 $\pm$ 0.09 &    0.40 $\pm$ 0.09 \\
NGC~1162     &  2.53 $\pm$ 0.01 &   1.12 $\pm$ 0.09 &    0.11 $\pm$ 0.09 &    0.30 $\pm$ 0.09 \\
NGC~1400     &  2.37 $\pm$ 0.01 &   0.65 $\pm$ 0.09 &    0.06 $\pm$ 0.09 &    0.17 $\pm$ 0.09 \\
NGC~1407     &  2.14 $\pm$ 0.01 &   1.02 $\pm$ 0.09 &   -0.27 $\pm$ 0.09 &    0.15 $\pm$ 0.09 \\
NGC~1600     &  2.32 $\pm$ 0.01 &   1.17 $\pm$ 0.09 &   -0.13 $\pm$ 0.09 &    0.33 $\pm$ 0.09 \\
NGC~1700     &  2.17 $\pm$ 0.01 &   0.57 $\pm$ 0.09 &    0.05 $\pm$ 0.09 &    0.20 $\pm$ 0.09 \\
NGC~2271     &  2.46 $\pm$ 0.01 &   1.04 $\pm$ 0.09 &    0.02 $\pm$ 0.09 &    0.35 $\pm$ 0.09 \\
NGC~2831     &  2.45 $\pm$ 0.02 &   1.17 $\pm$ 0.09 &   -0.40 $\pm$ 0.09 &    0.40 $\pm$ 0.09 \\
NGC~2832     &  2.26 $\pm$ 0.02 &   1.10 $\pm$ 0.09 &   -0.30 $\pm$ 0.09 &    0.25 $\pm$ 0.09 \\
NGC~2865     &  2.34 $\pm$ 0.01 &   1.10 $\pm$ 0.09 &    0.05 $\pm$ 0.09 &    0.40 $\pm$ 0.09 \\
NGC~3226     &  2.46 $\pm$ 0.01 &   1.10 $\pm$ 0.09 &   -0.15 $\pm$ 0.09 &    0.25 $\pm$ 0.09 \\
NGC~3377     &  2.43 $\pm$ 0.02 &   1.10 $\pm$ 0.09 &   -0.20 $\pm$ 0.09 &    0.28 $\pm$ 0.09 \\
NGC~3379     &  2.48 $\pm$ 0.02 &   1.12 $\pm$ 0.09 &   -0.15 $\pm$ 0.09 &    0.20 $\pm$ 0.09 \\
NGC~3384     &  2.30 $\pm$ 0.01 &   0.85 $\pm$ 0.09 &    0.00 $\pm$ 0.09 &    0.28 $\pm$ 0.09 \\
NGC~3557     &  2.42 $\pm$ 0.01 &   0.90 $\pm$ 0.09 &   -0.02 $\pm$ 0.09 &    0.30 $\pm$ 0.09 \\
NGC~3608     &  2.29 $\pm$ 0.01 &   1.17 $\pm$ 0.09 &   -0.25 $\pm$ 0.09 &    0.28 $\pm$ 0.09 \\
NGC~3640     &  2.37 $\pm$ 0.01 &   1.10 $\pm$ 0.09 &    0.40 $\pm$ 0.09 &    0.22 $\pm$ 0.09 \\
NGC~4240     &  2.25 $\pm$ 0.02 &   1.00 $\pm$ 0.09 &   -0.12 $\pm$ 0.09 &    0.02 $\pm$ 0.09 \\
NGC~4291     &  2.09 $\pm$ 0.01 &   1.10 $\pm$ 0.09 &   -0.20 $\pm$ 0.09 &   -0.07 $\pm$ 0.09 \\
NGC~4365     &  2.34 $\pm$ 0.01 &   1.05 $\pm$ 0.09 &   -0.02 $\pm$ 0.09 &    0.30 $\pm$ 0.09 \\
NGC~4374     &  2.43 $\pm$ 0.01 &   1.17 $\pm$ 0.09 &   -0.05 $\pm$ 0.09 &    0.20 $\pm$ 0.09 \\
NGC~4472     &  2.37 $\pm$ 0.01 &   1.00 $\pm$ 0.09 &    0.25 $\pm$ 0.09 &    0.15 $\pm$ 0.09 \\
NGC~4552     &  2.39 $\pm$ 0.01 &   1.00 $\pm$ 0.09 &    0.20 $\pm$ 0.09 &   -0.10 $\pm$ 0.09 \\
NGC~4636     &  2.33 $\pm$ 0.02 &   0.30 $\pm$ 0.09 &    0.28 $\pm$ 0.09 &   -0.05 $\pm$ 0.09 \\
NGC~4697     &  2.47 $\pm$ 0.01 &   1.17 $\pm$ 0.09 &    0.25 $\pm$ 0.09 &    0.20 $\pm$ 0.09 \\
NGC~5044     &  2.30 $\pm$ 0.01 &   1.10 $\pm$ 0.09 &   -0.10 $\pm$ 0.09 &    0.30 $\pm$ 0.09 \\	
NGC~5322     &  2.45 $\pm$ 0.01 &   1.10 $\pm$ 0.09 &    0.10 $\pm$ 0.09 &    0.25 $\pm$ 0.09 \\
Abell~754    &  2.30 $\pm$ 0.01 &   0.80 $\pm$ 0.09 &   -0.15 $\pm$ 0.09 &    0.30 $\pm$ 0.09 \\
Abell~970    &  2.31 $\pm$ 0.01 &   0.85 $\pm$ 0.09 &    0.09 $\pm$ 0.09 &    0.15 $\pm$ 0.09 \\
Abell~978    &  2.46 $\pm$ 0.01 &   0.90 $\pm$ 0.09 &   -0.02 $\pm$ 0.09 &    0.17 $\pm$ 0.09 \\
ESO~153-G003 &  2.40 $\pm$ 0.01 &   0.95 $\pm$ 0.09 &    0.12 $\pm$ 0.09 &    0.25 $\pm$ 0.09 \\
ESO~218-G002 &  2.49 $\pm$ 0.01 &   1.05 $\pm$ 0.09 &   -0.07 $\pm$ 0.09 &    0.22 $\pm$ 0.09 \\
ESO~318-G021 &  2.43 $\pm$ 0.01 &   1.17 $\pm$ 0.09 &   -0.12 $\pm$ 0.09 &    0.35 $\pm$ 0.09 \\
MCG-01-27-013&  2.38 $\pm$ 0.01 &   1.12 $\pm$ 0.09 &   -0.30 $\pm$ 0.09 &    0.30 $\pm$ 0.09 \\
MCG-02-13-009&  2.28 $\pm$ 0.01 &   0.97 $\pm$ 0.09 &   -0.32 $\pm$ 0.09 &    0.27 $\pm$ 0.09 \\
MCG-03-26-030&  2.36 $\pm$ 0.01 &   0.90 $\pm$ 0.09 &   -0.12 $\pm$ 0.09 &    0.17 $\pm$ 0.09 \\
\hline 
\hline
\end{tabular}
\end{center}
\caption[]{Mean values of stellar population parameters for our sample galaxies.}
\label{Pop_par3}
\end{table*}

\end{appendix}
\end{document}